\documentclass[aip,cha,reprint,a4paper,nofootinbib]{revtex4-1}
\setcitestyle{authoryear,round,sort,sectionbib}

\usepackage[utf8]{inputenc}

\usepackage[pdftex]{graphicx}
\usepackage{subfigure}
\usepackage{amsmath, amssymb}
\usepackage{enumitem}
\usepackage{color, colortbl}
\usepackage{afterpage}
\usepackage{url}
\usepackage{multirow}
\usepackage{rotating}
\usepackage{scalefnt}
\usepackage{hyperref}
\usepackage{ulem}
\usepackage{braket}
\usepackage{longtable}

\usepackage{soul}

\usepackage{braket,cleveref}

\newcommand{\footref}[1]{%
    $^{\ref{#1}}$%
}

\usepackage{xr}

\begin{document}

\title{The wisdom of the few: Predicting collective success from individual behavior}

\author{Manuel S. Mariani}
\email{manuel.mariani@business.uzh.ch}
\affiliation{Institute of Fundamental and Frontier Sciences, University of Electronic Science and Technology of China, Chengdu 610054, PR China}
\affiliation{URPP Social Networks, University of Zurich, CH-8050 Zurich, Switzerland}

\author{Yanina Gimenez} 
\affiliation{Core Data Science, Grandata, 550 15th Street, San Francisco, California 94103, USA}

\author{Jorge Brea}
\affiliation{Core Data Science, Grandata, 550 15th Street, San Francisco, California 94103, USA}

\author{Martin Minnoni}
\affiliation{Core Data Science, Grandata, 550 15th Street, San Francisco, California 94103, USA}

\author{René Algesheimer}
\affiliation{URPP Social Networks, University of Zurich, CH-8050 Zurich, Switzerland}

\author{Claudio J. Tessone}
\affiliation{URPP Social Networks, University of Zurich, CH-8050 Zurich, Switzerland}

\begin{abstract} 
Can we predict top-performing products, services, or businesses by only monitoring the behavior of a small set of individuals? 
Although most previous studies focused on the predictive power of "hub" individuals with many social contacts, which sources of customer behavioral data are needed to address this question remains unclear, mostly due to the scarcity of available datasets that simultaneously capture individuals' purchasing patterns and social interactions.
Here, we address this question in a unique, large-scale dataset that combines individuals' credit-card purchasing history with their social and mobility traits across an entire nation. Surprisingly, we find that the purchasing history alone enables the detection of small sets of ``discoverers" whose early purchases offer reliable success predictions for the brick-and-mortar stores they visit. In contrast with the assumptions by most existing studies on word-of-mouth processes, the hubs selected by social network centrality are not consistently predictive of success.
Our findings show that companies and organizations with access to large-scale purchasing data can detect the discoverers and leverage their behavior to anticipate market trends, without the need for social network data. 
\end{abstract}

\maketitle

\section{Introduction}

Measuring the potential of a new product, business, or service is a core challenge in management and marketing sciences~\citep{miklos2019collusion,muller2019effect}. Early-stage estimates are especially important because firms invest substantial resources into high-potential success stories (either their own or third-party ones), yet in social environments, success predictions are made difficult by social feedback mechanisms and serendipity~\citep{salganik2006experimental,van2014field}, and only a handful of products and services end up becoming highly-successful.
Can we early detect highly-successful products or businesses from the behavior of a small group of customers?
If individuals who forerun successful trends through their observed behavior are detectable, they could be beneficial not only for success forecasting, but also for seeding programs and the generation of new ideas for products and services. 
Theoretically, this possibility is backed up by long-standing theories of social influence and innovation diffusion, which suggest that a small minority of opinion leaders or influencers can have a disproportionate impact on the adoption of an innovation~\citep{katz1955personal,rogers2010diffusion}. To detect individuals whose adoptions signal increased odds of future success, extant studies have suggested two approaches.

A first approach would require to analyze social network data, under the assumption that individuals placed at favorable locations of the network have increased opportunities to contribute to products' or businesses' future growth~\citep{muller2019effect}. 
However, these efforts have mostly focused either on data from online social networks or on synthetic data from network diffusion models, and their results have been often inconsistent. Some studies have reported a consistent association between the social hubs' adoptions and future product success~\citep{goldenberg2009role,libai2013decomposing}, where the social hubs are defined as individuals with an outstanding number of social contacts. By contrast, other studies challenged this perspective by emphasizing that the contributions of social hubs to new product success is moderate~\citep{watts2007influentials}, and in online platforms where users reshare content, popularity predictions based on the network location of the seed user and early adopters are unreliable~\citep{bakshy2011everyone} and highly system-dependent~\citep{shulman2016predictability}. Empirical estimates of the predictive power of social hubs beyond online social networks are currently lacking, mostly because of the difficulty of matching social network data with transactional data.

A second approach would require to only analyze adoption data, under the assumption that we can detect from previous purchasing patterns ``harbinger" customers whose adoptions are predictive of failure or success~\citep{anderson2015harbingers}. Different groups of harbinger customers have been detected according to their ``flop rate", i.e., the proportion of purchased new products that failed within three years after launch. Although \cite{anderson2015harbingers} have shown that specific groups of harbinger customers can anticipate product failure or survival in the market, it remains unclear whether sets of harbinger customers detected from the purchasing history are better predictors of top-performing products and businesses than the widely-studied social hubs, which is a vital question for companies that need to decide how to analyze customer behavioral data to anticipate future market trends.

Here, we overcome the limitations of both streams of studies by showing that top-performing businesses can be early predicted by monitoring a small group of customers, without the need for social network data.
More specifically, we analyze an anonymized dataset\footnote{\label{data}All the data analyzed in this article are anonymized. The subjects of the analysis (individuals and stores) are represented by meaningless hashes in the dataset. All individuals are non-identifiable, all stores are nameless, and all transactions are innominate.} including individuals' time-stamped purchases in brick-and-mortar stores in an emerging country in America, and their social network based on mobile phone communication~\footref{data}. From this dataset, we extract individuals' purchases from a large-scale Credit Card Record (CCR) provided by a major bank (from June 2015 to May 2018); also, we collect individuals' communication and mobility patterns from a massive Call Data Record (CDR) provided by a mobile phone operator (from January to December 2016) operating in the same country as the bank. Importantly, we can match a substantial number of individuals\footnote{\label{individuals}Each individual in our dataset is \textit{non-identifiable}, meaning that (s)he is represented by a meaningless hash in the dataset, and there is no way to reconstruct the individuals' real identities.} in the CCR with the individuals in the CDR (see Online Appendix A). The availability of such rich dataset places us in a unique position to rigorously measure the predictive power of different sets of key individuals embedded in a large-scale market, and to evaluate the predictive power of sets of individuals detected from three different sources of data: purchasing, social network, and mobility data.

Our paper deepens our understanding of success predictability from individual behavior through three different (yet related) contributions.
First, from the Credit Card Record, we unveil the existence of \textit{discoverers}: a small set of individuals\footref{individuals} who repeatedly purchase in recently-opened stores that later become top-performing ones. 
Crucially, the discoverers' behavior cannot be explained by the number of stores they purchased in, which is revealed by comparing the discoverers' purchasing patterns against their expected behavior under a null model that preserves their overall activity level. The existence of discoverers in purchasing patterns from three different categories of stores indicates that the existence of customers who anticipate successful trends holds with a large degree of universality across different systems, ranging from online platforms where similar customers were found~\citep{medo2016identification} to the offline purchasing behavior analyzed here. We emphasize that the discoverers fundamentally differ from the harbinger customers studied by \cite{anderson2015harbingers}: the discoverers are a small set of $1\%$ (or $0.5\%$) customers who repeatedly purchase in recently-opened stores that later become top-performing ones (top $5\%$ or $10\%$), whereas the closest harbinger group detected by \cite{anderson2015harbingers} is composed of a set of $28\%$ low flop-rate customers who rarely purchase new products that disappear from the market within $3$ years after launch (which corresponds approximately to $60\%$ of all products).

Second, we demonstrate that the discoverers offer significantly more reliable out-of-sample predictions of store success than the widely-studied social hubs and other groups of top-individuals detected from social, purchasing, and mobility data. 
This finding deepens our understanding of success predictability from customer behavior: Our integration of purchase and social network data enables indeed the comparison of the predictive power of sets of top individuals detected from multiple sources of data, which allows us to connect with the social hubs' literature~\citep{watts2007influentials,goldenberg2009role,libai2013decomposing} and was missing in previous relevant studies~\citep{anderson2015harbingers,medo2016identification}. This comparison allows us to establish the superiority of the discoverers' predictive power compared to the social hubs, among others. Our findings complement previous work~\citep{anderson2015harbingers} which mostly focused on the predictability of the survival of new products, by showing that not only market survival can be predicted from the behavior of small sets of customers, but also the very top performers in the market.

Third, we measure the discoverers' socioeconomic traits, revealing their fundamental differences compared to the social hubs and store explorers (i.e., customers who purchase in many different stores), and the differences between the discoverers of different categories of stores.We find that the discoverer groups exhibit similar socioeconomic traits across the three categories of stores investigated here, but strikingly different demographic traits (age and gender). Among the socioeconomic traits, the relative difference between the discoverers' and all customers' median social-network centrality is statistically significant but relatively small, which suggests that in most cases, the discoverers' predictive power may not benefit from a disproportionate number of social contacts.

Our empirical findings support a new paradigm, which we call \textit{wisdom of the few}: to predict whether a new business will become a top-performing one, one can rely on the actions by a small set of individuals. Differently from commonly-held assumptions in management and network science, this set of predictive customers is not composed of the most socially-connected individuals. The detection of this set only requires the transaction history, enabling researchers and organizations to \textit{early predict top-performing businesses without the need for social network nor mobility data}.
The discoverer detection procedure and the predictive scheme developed here are general and can be applied for the prediction of the success of products, businesses, and services, provided that adoption data are available.

Our work has two managerial implications of immediate applicability. 
First, it formulates an unambiguous prediction problem to rigorously test claims about the predictive power of sets of customers. Second, many companies have access to purchasing data but may lack social network data for their customers: Banks have access to credit card transactions from/to individuals or businesses; owners of e-commerce websites possess purchasing history data; media-service providers have access to individuals' consumption records. 
Besides, offline and online retailers routinely collect customer transaction data through various kinds of loyalty cards. 
Remarkably, our work indicates that a single source of data -- the individuals' purchasing history -- is sufficient to detect the discoverers, and leverage their early adoptions to reliably forecast future market trends.

\section{Related literature}
\label{sec:literature}

Our article contributes to various streams of literature, including studies on success prediction, social hubs, and harbinger customers. 

Our results indicate that we can predict whether a new business will be a top-performing one by only monitoring the purchases by the small set of detected discoverers. Hence, the discoverers can be interpreted as a group of customers whose observed behavior exhibits a strong link with future collective outcomes.
Recently, moving beyond traditional approaches focused on modeling and forecasting the collective dynamics of customer behavior~\citep{bass1969new}, establishing a link between aggregate market behavior and the behavior of market segments or single individuals has gained increasing attention~\citep{peres2010innovation}.
This has benefited from the increasing availability of individual-level data, especially from online social networks, that include both social and consumption patterns. The main rationale is that market aggregate dynamics is the result of the adoptions or purchases by many individuals~\citep{peres2010innovation}; therefore, one might be able to predict aggregate market behavior from individual-level behavior, which is valuable to inform decisions for seeding strategies~\citep{muller2019effect}.

Aiming to link collective outcomes and individual-level behavior, there has been enormous interest in \textit{social hubs} -- namely, individuals with a large number of social contacts~\citep{goldenberg2009role} who are usually coveted targets for influencer marketing campaigns~\citep{lanz2019climb}. 
The roots of such interest on social hubs for adoption processes can be traced back to the well-documented social contagion and social learning mechanisms driving individuals' adoptions~\citep{tucker2008identifying,iyengar2011opinion,iyengar2015social,ma2015latent} and switching behavior~\citep{hu2019understanding}.
However, no consistent conclusions have been found on the social hubs’ predictive power. Some studies highlighted a consistent association between hubs’ adoptions and success~\citep{tucker2008identifying,goldenberg2009role,goldenberg2009zooming,libai2013decomposing,muller2019effect}. For example, by analyzing data from a large online social network, \cite{goldenberg2009role} found that it is possible to accurately distinguish between highly and moderately successful products by monitoring the hubs' early adoptions.
\cite{libai2013decomposing} focused on seeding strategies under a network diffusion model, finding that compared to random seeding strategy, seeding the social hubs results in a significant increase of diffusion success. 

Other studies emphasized that hubs can trigger large-scale cascades only on rare occasions and, thus, are not reliable predictors of success. This argument was first motivated by~\cite{watts2007influentials} via numerical simulations on synthetic networks under diffusion models.
Recent findings on observational data from various social media platforms support the idea that social hubs' content adoption might not be a reliable signal of future success in terms of content's reposts~\citep{cha2010measuring,bakshy2011everyone}, and the link between the structure of the social network around early content adopters and content future success might not generalize across different platforms~\citep{shulman2016predictability}.

Our work contributes to our understanding of the link between the social hubs' behavior and collective success by investigating the predictive power of the hubs' purchases in observational data from a large, offline market. This is notoriously hard to measure because of the difficulty to link social network data with transactional data; the availability of our unique dataset allows us to overcome this impasse.

Our work also complements recent works that aimed to detect harbingers of success or failure in purchase data~\citep{anderson2015harbingers}. 
\cite{anderson2015harbingers} defined four groups of harbinger customers based on the individuals' flop rate, defined as the fraction of purchased products that failed by disappearing from the market within two or three years after launch. Remarkably, the group of customers with the highest flop rate act as ``harbingers of failure", by
signaling reduced odds of success for the products they early adopt. This finding invalidates the conventional wisdom that early positive customer feedback is always good news for a firm. On the opposite hand, the harbinger group with the lowest flop rate (labeled as ``group 1" in \cite{anderson2015harbingers}) is the closest analog to the discoverers studied here.
Purchases by this group of customers are positively associated with the likelihood that a product will survive for more than three years~\citep{anderson2015harbingers}. 
The existence of different sets of harbinger customers supports the notion that not all positive feedback should be weighted the same, as specific sets of customers might foreshadow eventual success or failure.

Broadly speaking, our findings are supported by this paradigm; however, there are three fundamental differences between our work and \cite{anderson2015harbingers}.
First, the main success variable is different: \cite{anderson2015harbingers} mostly focused on the likelihood that a new product survives for more than three years after launch, leading to approximately $40\%$ "successful" new products. On the other hand, we focus on the likelihood that a new store ends up among the $10\%$ (or $5\%$) most popular ones, among same-category stores introduced in the same month. Therefore, our definition of success focuses on the top-performing businesses, which is a much stricter requirement than market survival. Besides, as our store-level success variable only compares a store with other stores opened in the same month, it can be evaluated over a substantially shorter time window compared to the $2$ or $3$ years necessary to assess whether a product has failed or not~\citep{anderson2015harbingers}, which is a practical advantage for leveraging the discoverers' predictive power in real-world forecasting applications. Third, and most importantly, we have access to social network data, which allows us to both compare the discoverers' predictive power against those by other groups of top-individuals (including the widely-studied social hubs), and to measure the discoverers' social-network centrality.

\begin{table*}[t]
\begin{center}
\begin{tabular}{ |c|c|c|c|c|c|c|c| }
\hline
  Category & $T$ & $S_{tot}$ & $S_{tr}$ & $S_{val}$ & $\braket{c}$ & $I$ & $\braket{k}$ \\
  \hline 
  Eating places & $9,946,197$ & $115,383$  & $28,196$ & $11,654$ & $154.44$ & $1,323,647$ & $7.51$\\
  \hline 
  Food stores & $13,606,954$ & $97,096$ & $15,046$ & $6,456$ & $240.26$ & $2,061,374$ & $6.60$ \\
  \hline
  Clothing stores & $3,967,272$ & $66,753$ & $12,809$ & $3,822$ & $92.80$ & $1,166,701$ & $3.40$ \\
\hline
\end{tabular}
\end{center}
\caption{\textbf{Basic data properties.} We report basic properties of the (nameless) eating places, food stores, and clothing stores found in the Credit Card Record (see Online Appendix A).
The variables represent: the total number of first-time transactions in stores, $T$, over the training period; the total number of stores that received transactions, $S_{tot}$; the number of stores that received their first transaction within the training period (excluding the last two months), $S_{tr}$; the number of stores that received their first transaction within the validation period (excluding the last two months), $S_{val}$; the average number of first-time customers per store, $\braket{c}$, over the training period; the number of individuals who made purchases during the training period (excluding the last two months), $I$; the average number of stores visited by the individual during the training period (excluding the last two months), $\braket{k}$. }
\label{table}
\end{table*}

\begin{figure*}[t]
        \centering
        \includegraphics[scale=0.35]{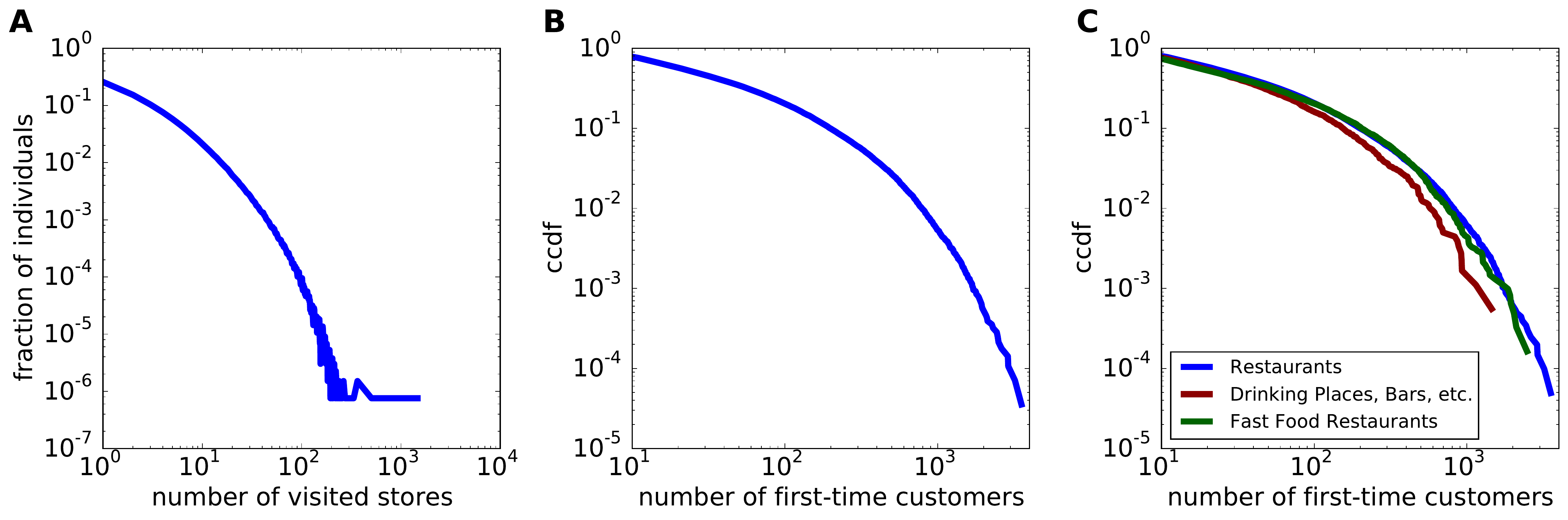}
        \caption{\textbf{Individuals' activity patterns for eating places. (A)} Distribution of the number of visited stores per individual. \textbf{(B)} Complementary cumulative distribution function of the number of first-time customers per store. \textbf{(C)} Same as (B) for each Merchant Category Code (see Table S1).}
        \label{fig:fig1}
    \end{figure*}

Overall, the predictive problem studied here fundamentally differs from the harbinger customer and the predictive problem studied by~\cite{anderson2015harbingers}, and we are additionally able to integrate transactional data with social network data, which enables a detailed comparison of the discoverers' and the social hubs' predictive power.
From a success prediction standpoint, our results complement their findings, by demonstrating that companies can leverage individual-level purchasing patterns not only to predict which new products will survive or fail, but also which new stores will end up among the top-performing ones. 

Following up on the work by~\cite{anderson2015harbingers}, \cite{simester2019surprising} detected and characterized "harbinger zip codes", i.e., zip codes whose households' purchases of a new products signal increased odds of failure. Their characterization resorted to the integration of multiple sources of data, including transactional data, coupon data, and demographic data from a mass merchandise store; transactional data from a private label apparel retailer, data on contributions to congressional election candidates and election outcomes; data on house prices. This data-rich setting allowed \cite{simester2019surprising} to demonstrate that the harbinger effect holds at the zip code level, and that households located in harbinger zip codes make decisions that differ from those in neighboring zip codes across a wide variety of decision contexts, beyond purchase decisions. \cite{simester2019surprising}'s findings lead us to conjecture that the existence and predictive power of the discoverers likely extend far beyond the retail context investigated here, and might be found at different scales of investigation, beyond the individual-customer scale studied here. We note that \cite{simester2019surprising} did not integrate social network data in their analysis, which leaves it unclear whether various sets of harbinger households are located at (un)favorable locations in the social network.

\section{Data}
\label{sec:data}

We analyze a unique anonymized dataset\footref{data} that includes credit-card transactions of (non-identifiable) individuals\footref{individuals} over a three-year temporal window (from June 2015 to May 2018). By recording only the first purchase of each individual\footref{individuals} in each nameless store and after pre-filtering the data (see Online Appendix A), the complete dataset includes more than $140$ million of time-stamped transactions\footref{data}.
In the following, whenever we will refer to an individual, this should be interpreted as a non-identifiable individual whose real identity is impossible to identify. Accordingly, whenever we will refer to a group of ``detected" individuals, this should be interpreted as a group of individuals whose real identities cannot be identified. For the sake of better readability, when referring to the analyzed datasets and the individuals in the following, we shall omit the labels ``anonymized" and ``non-identifiable".

The first observation is that the dataset is highly heterogeneous, encompassing categories as diverse as book stores, tech stores, and florists, among many others.
To control for this heterogeneity, we restrict our analysis to three categories of stores that have a well-defined interpretation: eating places, clothing stores, and food stores. The stores that belong to each of the three categories are selected based on their Merchant Category Code (MCC) information present in our database -- we refer to Online Appendix A for details. 

We split the dataset's time span into three periods: a $6$-month pre-filtering period that is used to determine which stores appear for the first time in the training period that follows; a $18$-month training period that ranges from December 2015 to May 2017, where we aim to detect groups of key individuals (described below); a $12$-month validation period that is used to perform and validate success predictions. 
The main rationale behind our choice of the relative duration of training and validation period is that the validation period should be long enough to include a substantial number of new stores to validate our predictions, yet short enough to not exceedingly restrict the training period where we detect the key individuals.
Our predictive results are robust with respect to small variations of the relative duration of training and validation period, as discussed in Online Appendix E.

 Table~\ref{table} summarizes basic data properties. 
The individuals' number of visited stores, $k$, is highly heterogeneous, with a small number of outliers with a large number of visited stores (Fig.~\ref{fig:fig1}A and Online Appendix F). Similarly, the number of first-time customers per store, $c$, is highly heterogeneous (Figs.~\ref{fig:fig1}B) and dependent on the store's MCC category (Figs.~\ref{fig:fig1}C and Online Appendix F). 
A given store's number of first-time customers represents its ability to acquire new customers~\citep{bell2017offline}, and we use it to operationalize store success.
Besides the CCR, we analyze a CDR from the same market over a one-year period that overlaps with the CCR's time span. For a subset of the population, we know both the social behavior (in terms of mobile phone communication) and the economic behavior (in terms of monetary purchases in stores) -- see Online Appendix B for details. 
From the CDR, we extract individual-level traits related to their centrality in the social network (time-averaged degree~\citep{goldenberg2009role}, collective influence~\citep{morone2015influence}, and social diversity~\citep{eagle2010network}), and mobility (radius of gyration~\citep{gonzalez2008understanding} and mobility diversity~\citep{song2010limits}). We refer to Online Appendices B--C for the details of the measured features.

\section{Results}
\label{sec:results}

Motivated by our predictive question outlined in the Introduction, we adopt a statistical procedure that seeks to find individuals -- referred to as \textit{discoverers} -- who are persistently able to discover stores with a high potential of becoming popular. We then measure the out-of-sample predictive power of the discoverers, and compare it against the predictive power of seven other groups of top individuals detected from the purchase history, social network, and mobility data. Finally, we provide a demographic and socioeconomic characterization of the discoverers, social hubs, and store explorers.

\subsection{Discoverer detection}

\begin{figure*}[t]
        \centering
        \includegraphics[scale=0.35]{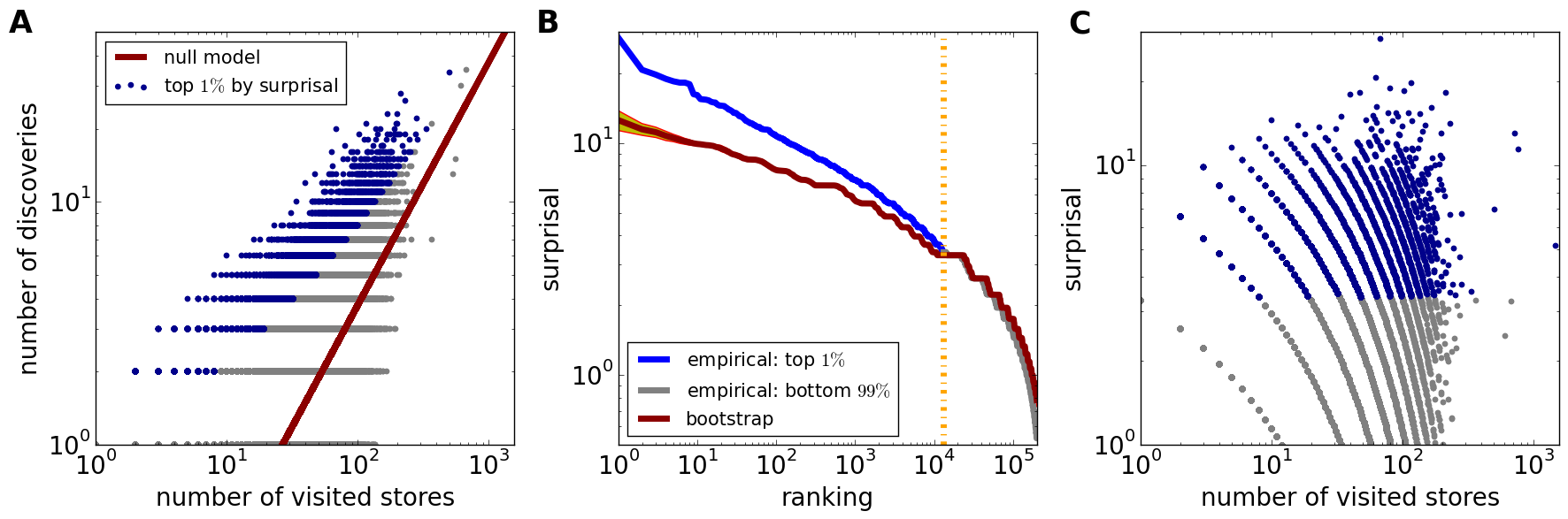}
        \caption{\textbf{Individuals' discovery patterns for eating places. (A)} Number of discoveries vs. number of visited stores. The two variables are positively correlated, yet the trend deviates from the expected one under the null model, and some individuals collect significantly more discoveries than expected from their activity alone. The straight line represents the expectation under the null model, whereas the blue dots mark the top-$1\%$ individuals by surprisal. \textbf{(B)} Zipf plot of the individuals' surprisal for the empirical data and the null model. The discrepancy between the two curves is significant, meaning that the surprisal values of most of the top-$1\%$ individuals by empirical surprisal (i.e., those at the left of the vertical dashed line) are significantly above the surprisal values expected for their ranking position. \textbf{(C)} Surprisal vs. number of visited stores. The two variables are only weakly correlated. }
        \label{fig:fig1disc}
    \end{figure*}

To detect the discoverers, inspired by a method previously applied to e-commerce data~\citep{medo2016identification}, we define a \textit{discovery} as an early purchase in a store that later becomes successful. Throughout this article, to define a discovery, we consider a purchase as an early purchase if it happened no later than $90$ days after the store received its first transaction; we consider a store as successful if it ends up among the top $10\%$ by number of first-time customers among stores introduced in the same month and of the same MCC. The discoverers' predictive power is reasonably robust with respect to alternative specifications of the early purchase window and success, as discussed in Online Appendices D--E. 

The discoverers are selected by a measure of statistical unexpectedness -- called \textit{surprisal} -- that quantifies how unlikely an individual's observed number of discoveries was under a null model that preserves the individuals' level of activity (in terms of number of visited stores) and assumes that everyone has the same likelihood to collect a discovery -- see Appendix~\ref{sec:discoverers} for details. To ensure that high surprisal values are not due to high levels of activity, we perform a bootstrap procedure by resampling the individuals' number of discoveries from the null model distribution (preserving the individuals' number of visited stores), and comparing the empirical largest surprisal scores against the largest scores from the bootstrap -- see Appendix~\ref{sec:bootstrap} for details. In general, it is not guaranteed that there exists a sizeable set of individuals who achieve not only a number of discoveries that significantly deviate from the expectations from the null model, but also a surprisal score that is significantly larger than the largest surprisal scores observed in the bootstrap procedure.

However, we find that while the number of discoveries per individual is positively correlated with the number of visited stores (Fig.~\ref{fig:fig1disc}A), the deviations from the trend are significant and cannot be explained by chance. 
To rule out the possibility that high values of surprisal are obtained through random fluctuations, we compare the empirical surprisal values of the detected discoverers with the top-surprisal values observed by resampling the individuals' number of discoveries from the null-model distribution (see Appendix~\ref{sec:bootstrap}).
We find that the largest empirical surprisal values are significantly larger than the largest surprisal values obtained by resampling the individuals' number of discoveries (Fig.~\ref{fig:fig1disc}B). The surprisal values are weakly correlated with the individuals' number of visited stores, indicating that activity alone is not a good proxy for an individual's propensity to discover successful stores (Fig.~\ref{fig:fig1disc}C). The results in Fig.~\ref{fig:fig1disc} refer to eating places; results for other store categories are qualitatively similar (Online Appendix F). Taken together, these findings indicate that some individuals exhibit a clear propensity to purchase in recently-opened stores that later become successful, and it is highly unlikely that this pattern can be explained solely by the individuals' level of activity.

\subsection{Tracking detected individuals to predict store success}

\begin{figure*}[t]
        \centering
        \includegraphics[scale=0.25]{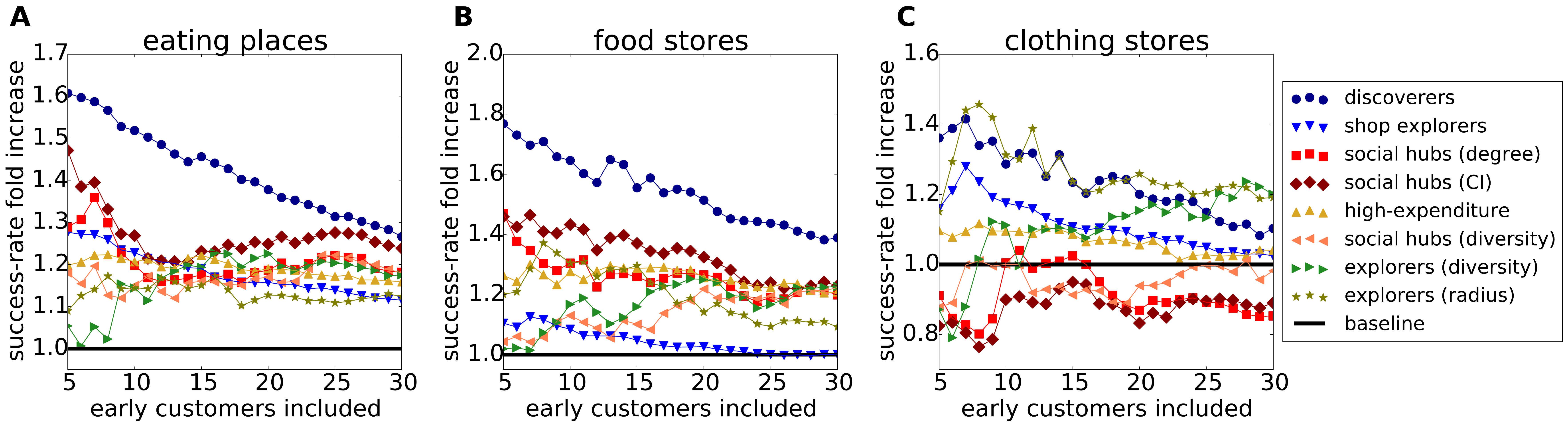}
        \caption{\textbf{Wisdom of the few: Success predictions based on groups of key individuals.} Success-rate fold increase for stores visited early on by different groups of top-individuals, for eating places \textbf{(A)}, food stores \textbf{(B)}, and clothing stores \textbf{(C)} that received their first transaction within the validation period. 
        The discoverers' success rate is substantially larger than that of other groups for eating and food stores. The discoverers and the explorers selected by radius of gyration are similarly competitive for clothing stores, and they outperform the other groups.
        Compared to the discoverers, the performance by other groups of selected individuals is inconsistent or weaker. For example, the social hubs are associated with an above-baseline success rate for eating places and food stores, but not for clothing stores. The explorers (selected by radius of gyration) exhibit a success rate comparable to the discoverers' one for clothing stores, but their predictive power is substantially weaker for eating places and food stores.
        }
         \label{fig:fig2}
    \end{figure*}

The previous analysis reveals that there exist individuals who repeatedly purchase in recent stores that later end up being successful. Yet, the detected individuals would have true predictive value only if they would be predictive of success over a ``validation period" that follows the training period within which they were detected. 
Is the discoverers' tendency to collect discoveries persistent enough over time to allow us to track their actions for reliable out-of-sample success predictions? This is not obvious \textit{a priori}, but if it would be the case, it would suggest that discovering successful stores is a persistent behavioral trait of the discoverers.
Besides, are the predictions made by tracking the discoverers' purchases more accurate than those obtained through other small groups of top individuals?

To answer these questions, we analyze stores that received their first transaction within the validation period (from June 2017 to May 2018) to evaluate the predictive power of top-individuals detected within the training period (i.e., by only using data from December 2015 to May 2017). We compare the discoverers' predictive power against groups of top individuals selected by network centrality measures (the aforementioned \textit{hubs} selected by degree~\citep{goldenberg2009role}, collective influence~\citep{morone2015influence}, and social diversity~\citep{eagle2010network}), total expenditures (\textit{high-expenditure} individuals~\citep{di2018sequences}), number of visited stores (\textit{store explorers}), and mobility-related features (\textit{explorers} selected by mobility diversity~\citep{song2010limits} and radius of gyration~\citep{gonzalez2008understanding}). Comparing the discoverers' performance against that by seven alternative groups of top-individuals allows us to ensure that the magnitude and cross-category consistency of the discoverers' predictive power cannot be matched by other known customer social, economic, and mobility traits. As for the social hubs, the inclusion of the hubs selected by collective influence is motivated by recent network-science findings that indicate that sets of individuals with high collective influence significantly outperform sets of high-degree individuals in triggering large-scale diffusion processes~\citep{morone2015influence,lu2016vital}.
We refer to the Online Appendices B--C for all details on the detection of the groups of top-individuals included in the analysis.

We consider the classification problem where we aim to predict whether a store introduced in the validation period will be among the top-$10\%$ shops by final number of first-time customers, among the stores with the same MCC that received their first transaction in the same month -- if this is the case, we say that the store is \textit{successful}. Our definition of success factors out two potential confounding factors in our measure of success: store age and category (see Online Appendix F). 
To quantify the predictive power of a given group of individuals, $\mathcal{I}$, we measure the fraction of successful stores among those that featured an individual in $\mathcal{I}$ among the earliest $w$ first-time customers -- we refer to this fraction as $\mathcal{I}$'s \textit{success rate}. We divide this success rate by the baseline success rate given by the fraction of successful stores among those that received at least $w$ first-time customers by the end of the validation period, obtaining the fold increase of $\mathcal{I}$'s success rate with respect to the baseline expectation. For the sake of brevity, we refer to this ratio as $\mathcal{I}$'s \textit{success-rate fold increase}. In other words, for each group $\mathcal{I}$, we are defining a Naive Bayes Classifier that classifies a store as successful if an individual in $\mathcal{I}$ is found among the earliest $w$ customers, unsuccessful otherwise~\citep{sarigol2014predicting}. According to this interpretation, $\mathcal{I}$'s success rate can be interpreted as the precision of $\mathcal{I}$'s classifier, and standard classifier evaluation metrics such as recall and the Matthews' correlation coefficient can be evaluated~\citep{powers2011evaluation} -- see Online Appendices E-F for the complete predictive results.

We find that the detected discoverers exhibit a consistent predictive signal across all three store categories (Fig.~\ref{fig:fig2}).
In particular, for eating places and food stores, the discoverers exhibit the largest success-rate fold increase for all numbers of included early customers (Figs.~\ref{fig:fig2}A--B). For clothing stores, explorers selected by radius of gyration exhibit a similar success rate to the discoverers' one (Figs.~\ref{fig:fig2}C). The early purchases by other classes of individuals might still be associated with larger-than-baseline success rates, yet none of the other groups of individuals is competitive across all three store categories. Based on existing literature, the social hubs are the most interesting group to compare against the discoverers. When considering the earliest $10$ first-time customers of eating places, the social hubs selected by degree and collective influence achieve an above-baseline success rate (1.20 and 1.27, respectively), even though smaller than that achieved by the discoverers (1.52). However, the same does not hold for clothing stores: the stores that received a purchase by a social hub (selected by degree) exhibit a success rate that is comparable to the baseline ($1.00$ fold increase), whereas the discoverers still exhibit a significant predictive power ($1.29$ success-rate fold increase).
We refer to Fig.~\ref{fig:fig2} for the full results.

Similar conclusions can be drawn from the results obtained with two more prediction evaluation metrics: the Matthews' correlation coefficient\footnote{Recent findings indicate that to evaluate the overall performance of a classifier for a classification problem with an imbalanced set, the Matthews' correlation coefficient should be preferred over the more traditional F1 score and accuracy~\citep{chicco2020advantages}.} and the positive likelihood ratio~\citep{powers2011evaluation} -- we refer to Online Appendices D--E for the detailed results.
The recall metric (namely, the fraction of successful stores that are classified as successful) exhibits a different trend compared to the success rate because by construction, it favors groups of individuals that purchased in many different stores. Because of this, store explorers exhibit the largest recall values across all three categories, followed by the discoverers and high-expenditure individuals (see Online Appendix D).
The results for the recall metric indicate that while early purchases by the discoverers are predictive of success, there exist stores that succeed without an early visit by the discoverers. For example, out of $10,301$ eating places introduced in the validation period that received at least 10 customers, $1,172$ are successful, and $38.8\%$ of them received a discoverer among the earliest $10$ first-time customers.
Therefore, not all successful stores received an early purchase by a discoverer. Still, an eating place that received a discoverer among the earliest 10 customers is $62.5\%$ more likely to be successful than one that did not. Similar conclusions can be drawn by considering different numbers of early customers and store categories -- See Online Appendix E.

Beyond the classification problem, one can also investigate whether stores that received a discoverer among the earliest customers tend to receive a larger number of customers in the future. We find that this is the case across all three store categories. For example, among the eating places that received their first transaction in the validation period, those that had a discoverer among the earliest $10$ customers tend to gain approximately $40.3\%$ more first-time customers that the average store with the same age and category. Similar results are obtained for food and clothing stores, where the relative fold increases of the final number of customers associated with the presence of a discoverer among the earliest customers are $33.2\%$ and $17.0\%$, respectively (see Online Appendix F for detailed results).

Overall, our results are reasonably robust with respect to variations in the parameters of the analysis, including the discoverer detection parameters, the threshold used to define successful stores, the relative duration of training and validation periods, the fraction of individuals included in each group (see Online Appendices D--E for a detailed presentation).
We also notice that it is possible to consistently improve the predictions' success rates by pairing the discoverers with other groups of individuals, yet these improvements tend to be marginal (see Online Appendix E).
Taken together, our results indicate that early purchases by the discoverers are typically associated with increased odds of future success and an increased number of future customers with respect to the purchases by other groups of top-individuals. By contrast, the early-purchases by the hubs are not reliable predictors of success for the visited store.

 \begin{figure*}[t]
        \centering
        \includegraphics[scale=0.09]{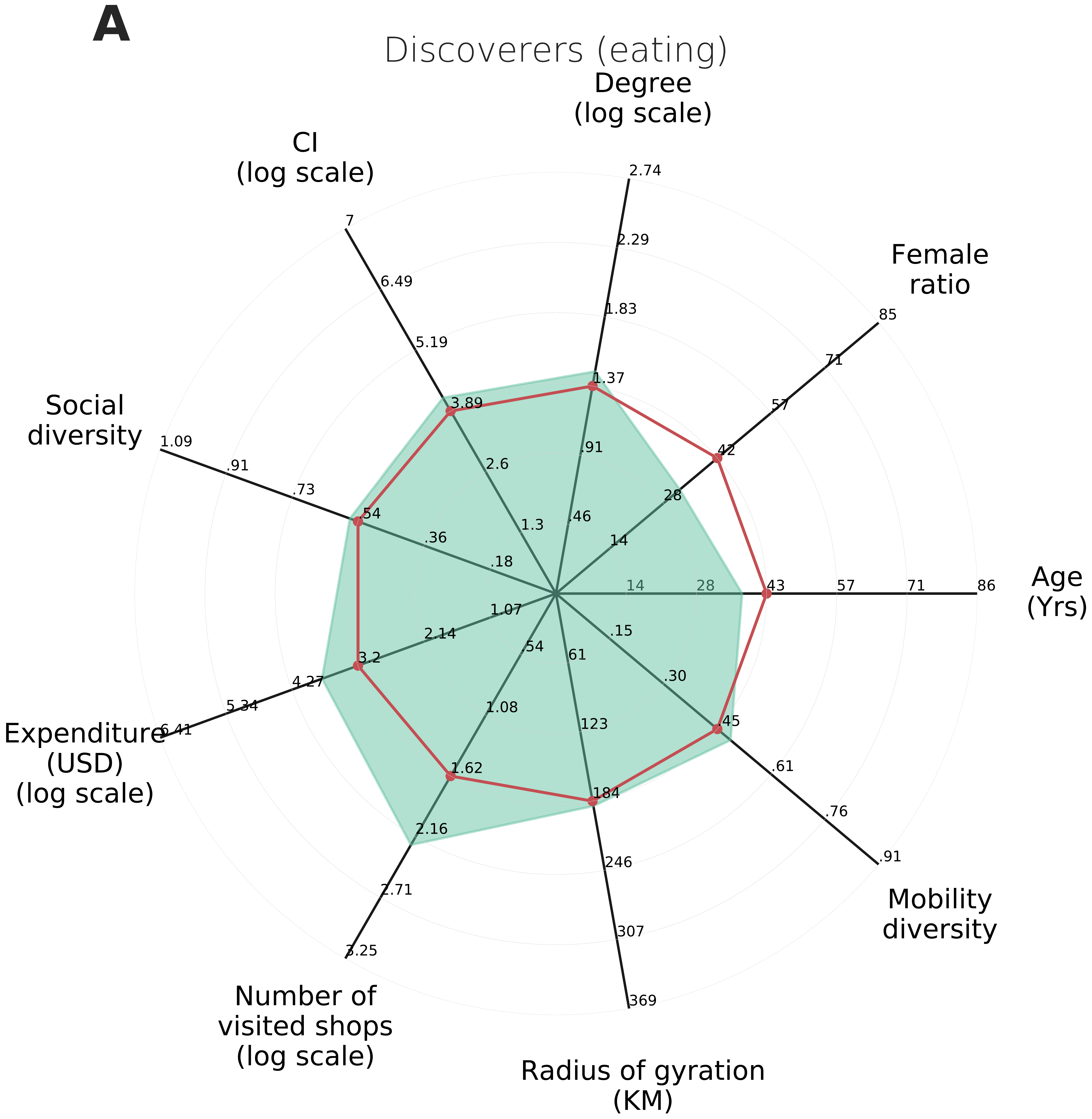}
        \includegraphics[scale=0.3]{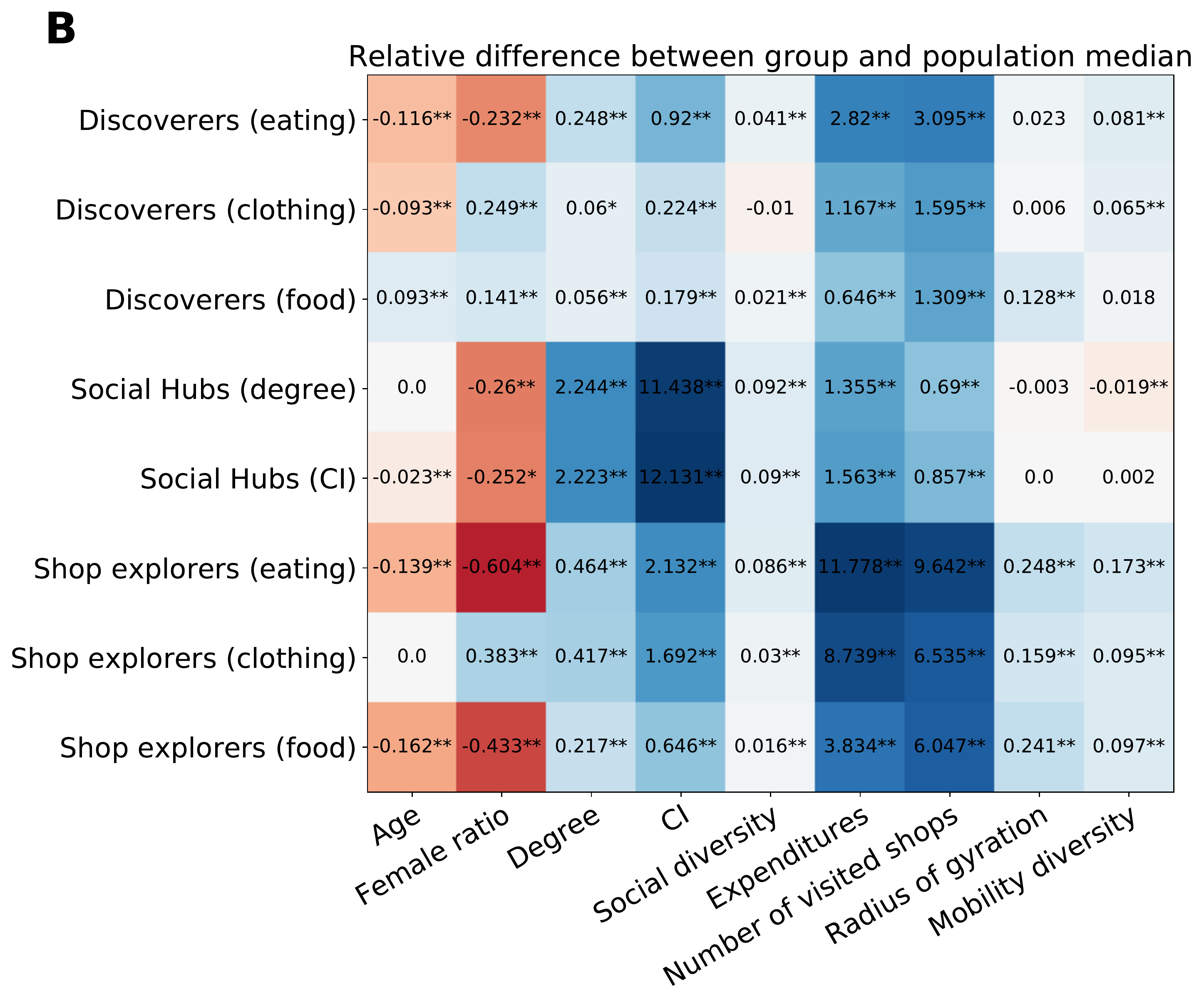}
        \caption{\textbf{Profiling key individuals: Discoverers, social hubs, and explorers.} \textbf{(A)} The radar plot illustrates the profiling of a group of individuals (discoverers of eating places). We compare the individuals' traits (represented by the radars' axes) with the traits' population medians (connected by the red polygon). \textbf{(B)} For each group of individuals (row), $\mathcal{I}$, and each trait (column), $\mathcal{T}$, we report the difference between $\mathcal{T}$'s median for individuals in $\mathcal{I}$ and the population median, normalized by the population median. Some of the discoverers' socioeconomic and mobility traits are consistent: regardless of store category, the typical discoverer exhibits above-median total expenditures, number of visited stores, collective influence, number of social contacts, and mobility diversity. By contrast, the discoverers' demographic traits (age and gender) are category-dependent. The traits of social hubs and store explorers are also reported. Alongside the reported relative differences, single ($^{*}$) or double ($^{**}$) stars mark significance levels below $P<0.05$ and $P<0.01$, respectively, under the 
        binomial test for female ratio, and the Mood's median test for all other features (see Appendix~\ref{sec:statistical}).} 
         \label{fig:fig4}
    \end{figure*}

\subsection{Socioeconomic characterization of discoverers, social hubs, and explorers}

Having detected the discoverers and measured their predictive power, it is inevitable to investigate which traits make them different from ordinary individuals and social hubs.
This is relevant for companies in order to detect prospective discoverers in scenarios where transaction data for their customers are not available, and potentially nudge into behaving as discoverers individuals who exhibit appropriate combinations of traits. We start by observing that there is little overlap between pairs of groups of discoverers of different categories of stores: The largest overlap is $380$ common discoverers among the discoverers of eating places and food stores (corresponding to the $2.8\%$ of the discoverers of eating places and the $1.8\%$ of the discoverers of food stores); the largest Jaccard similarity is $0.015$ observed between the sets of discoverers of eating places and clothing stores.
This indicates that it is unlikely that a discoverer in one category is also a discoverer in another category of stores, suggesting that the discoverers tend to be highly specialized.

Interesting demographic differences emerge across the groups of individuals (Fig.~\ref{fig:fig4}). 
The discoverers' demographic traits are consistent with the store explorers' ones for eating places (the members of both groups tend to be males whose age is below the population median) and for clothing stores (the members of both groups tend to be female). A clear difference emerges for food stores: the store explorers tend to be males with below-median age, yet the discoverers tend to be female with above-median age. Hence, not only store explorers are not necessarily discoverers (Fig.~\ref{fig:fig1disc}C), but the two groups of individuals can also exhibit starkly different demographic traits.

As for socioeconomic traits, across all three store categories, the trait where the discoverers' median deviates the most from the population median is the number of visited stores, followed by total expenditures, collective influence, and number of contacts (for all these differences, $P<0.01$ according to the Mood's median test). The discoverers' large expenditures and number of social contacts suggest that they have a higher socioeconomic status and degree of social connectedness than ordinary individuals. In this sense, they respect the traits of early adopters outlined in Rogers' seminal theory on the diffusion of innovations~\citep{rogers2010diffusion}.
Yet, the discoverers are not outstanding in any of these traits: for example, the social hubs selected by degree exhibit a substantially larger number of contacts, and store explorers purchase in a substantially larger number of stores. We also notice that the discoverers' mobility diversity is slightly above the population median, with the smallest relative difference observed for the discoverers of food stores ($+2.0\%, P<0.05$).
 Overall, these results support a scenario where the discoverers benefit from various socioeconomic traits, yet none of the investigated traits is predictive of the behavior of an individual as a discoverer. Further research is necessary to assess whether alternative socioeconomic traits, or some psychological traits, can be leveraged to accurately distinguish the discoverers from the rest of the population.

Intriguingly, we can compare the socioeconomic and demographic traits of the social hubs with results obtained by previous studies.
We find that the social hubs (selected by degree and collective influence) exhibit above-median expenditures ($P<0.01$).
This is in qualitative agreement with previous studies that reported that expenditures are a reliable proxy for monthly income~\citep{di2018sequences}, and social hubs tend to have a higher economic status~\citep{luo2017inferring}. The social hubs also exhibit an above-median number of visited stores, but not an above-median radius of gyration nor mobility diversity. In line with previous findings on online platforms~\citep{goldenberg2009role}, social hubs tend to be males.
 Compared to the discoverers, the social hubs exhibit a smaller radius of gyration and a smaller number of visited stores. Store explorers tend to be more central in the social network and to travel for a longer distance (as measured by their radius of gyration) than the discoverers.

 \section{Discussion}
 \label{sec:discussion}
 
The canonical narrative in management and network sciences is that influential individuals can be detected from their position in social networks, and the social hubs found in social network data are associated with a stronger future growth of their adopted products or services~\citep{muller2019effect}. Our results question this paradigm by revealing that if our goal is to predict success, early purchases by social hubs are only predictive of success for specific store categories, and their predictive signal is weaker than that observed for discoverers. This suggests that consumer heterogeneity is a more important driver of stores' success than word-of-mouth processes~\citep{peres2010innovation}.
By contrast, the discoverers are not among the most central individuals in the social network that we analyzed, yet their early monetary transactions are predictive of the stores' potential, without the need to explicitly consider the social network the discoverers are embedded in. Therefore, companies and organizations that have only access to transaction data would be able to use our approach to detect key individuals for success prediction. 
 
These conclusions were reached by integrating massive purchasing data and social network data, which allowed us to compare the predictive power of the discoverers against three groups of social hubs. Hence, our work links the recent literature on ``predictive customers" detected from purchasing data~\citep{anderson2015harbingers,medo2016identification,simester2019surprising} with the well-established literature on the role of social hubs in adoption processes~\citep{watts2007influentials,goldenberg2009role,goldenberg2009zooming,libai2013decomposing,muller2019effect}. Compared to the recently-studied ``harbinger customers"~\citep{anderson2015harbingers} who rarely purchase new products that later disappear from the market, the discoverers repeatedly early purchase in \textit{top-performing} business, allowing for consistent out-of-sample predictions of new top-performers.

Data-driven success predictions have recently gained traction in various scientific disciplines beyond management science, including the science of science~\citep{fortunato2018science}, social sciences~\citep{hofman2017prediction}, and macroeconomics~\citep{tacchella2018dynamical}, among others.
The predictive problem framed here and the discoverer detection procedure hold promise for application to any human activity data where individual units make adoption decisions in social environment. Future studies might indeed search for discoverers among online users downloading or resharing content in social media, organizations adopting new technologies, and scientists choosing research topics, among others. From a management research standpoint, future studies could improve our predictions by integrating the discoverers' predictive signal with other success predictors, including new store preannouncement activities~\citep{rao2019new} and geographic information~\citep{simester2019surprising}.

We conclude by highlighting two limitations of our study, which open exciting research avenues.
First, we implicitly assumed that mobile-phone communication data provide us with sufficiently good estimates of the individuals' centrality in the actual network of social contacts. On the other hand, with the rise of online social networks and instant messaging platforms, the phone communication network only provides us with a partial representation of the actual communication flows in society. Obtaining a complete representation of the social communication patterns across an entire nation is clearly unattainable.
Therefore, while the incompleteness of our social graph is a limitation of our work, it also mimics a real-world managerial scenario where an organization has only access to incomplete social information about its customers. Future studies need to generalize our results to online communities or virtual worlds where data on individuals' social connections, communication, and behavior are (nearly) complete.

Second, based on our data, we cannot establish any causal connection between the discoverers' purchases and the success of the stores. We have addressed the predictive question of whether the purchases by selected key individuals are consistent indicators of future success, but we did not address the question of whether the detected discoverers play an active or passive role in the market dynamics. In other words, it remains open to assess whether the discoverers can actively accelerate the stores' growth (similarly as the long-studied opinion leaders and influencers are assumed to do~\citep{muller2019effect}): if this is the case (at least for some of them), due to their lower centrality compared to the social hubs, they could be cost-effective targets for seeding campaigns. Another open question is whether delaying the discoverers' purchases would slow down the stores' growth (similarly as reported in a recent experiment for natural early adopters of a new technology~\citep{catalini2017early}). In future research, we will examine these possibilities through field experiments.
As the discoverers effectively anticipate future market trends by manifesting their present needs through their purchases, they can also serve to generate effective ideas for new products and services in a similar way as lead customers do~\citep{lilien2002performance}.

Not all early customers are the same. In large-scale markets, there exists a reliable predictive power hidden in the actions of small sets of individuals, which can be unveiled through appropriate statistical methods applied to the purchasing history. Deepening our understanding of the mechanisms, motivations, and personal values behind the emergence of such predictive customers is a fascinating challenge for future research.

\appendix

 \section{Discoverer detection}
 \label{sec:discoverers}
 
 We aim to quantify individuals' tendency to purchase in recent stores that later become successful. To this end, we define a \textit{discovery} as an event where an individual purchases in a store no later than $\Delta$ days after the store received its first transaction, and the store turns out to be successful.
A store is considered as \textit{successful} if it is among the top-$z\%$ stores by final number of first-time customers (at the end of the training period), among the stores with the same Merchant Category Code that received their first transaction in the same month. Stores that received their first transaction within the last two months of the training period are excluded from the analysis; similarly, transactions in that period do not contribute to the individuals' number of visited stores and discoveries, but only to the stores' number of customers. In the following, we denote the individuals who made at least one transaction in a store included in the analysis as $i\in\{1,\dots,I\}$, where $I$ is the total number of active individuals.

We determine individual $i$'s propensity to discover successful stores as the unexpectedness of $i$'s observed number of discoveries, $d_i^*$, in terms of the statistical surprisal $s_i(d_i^*)=-\log{P_i(d\geq d_i^*)}$, where $P_i(d\geq d_i^*)$ represents the probability that individual $i$ collected $d_i^*$ or more discoveries given its total number of visited stores, $k_i$. More specifically, the probability $P_i(d\geq d_i^*)$ is determined analytically as follows.
We consider a process where each individual $i$ draws (without replacement) $k_i$ marbles out of an urn which contains $T$ marbles, $D$ of which are labelled as ``discoveries"; $T$ and $D$ match the empirical number of first-time transactions and discoveries, respectively: $T=\sum_i k_i$, and $D=\sum_i d_i^*$. This choice of the null model aims to factor out individuals' level of activity, $k_i$, from the surprisal measure. The probability that $i$ achieves $d$ discoveries follows the hypergeometric distribution with mean $\braket{d_i}=k_i\,D/T$:
\begin{equation}
P_i(d|k_i)=\frac{ \binom{D}{d}\binom{T-D}{k_i-d}}{{\binom{T}{k_i}}}.
\label{hypergeo}
\end{equation}
Individual $i$'s surprisal is given by
\begin{equation}
s_i=-\log{P_i(d\geq d_i^*)}=-\log{\Biggl(\sum_{d=d_i^*}^{k_i} P_i(d|k_i)\Biggr)}.
\label{surprisal}
\end{equation}
The discoverers are detected as the top-$\tau\%$ individuals by $s$. Results in the main text have been obtained with $z=10\%$, $\Delta=90$ days, $\tau=1\%$; results for different values of $z,\Delta,\tau$ are reported in Figs. S7--S9. 

The surprisal metric naturally rewards individuals whose observed number of discoveries $d_i^*$ far exceeds their expected number of discoveries under the null model.
For example, for eating places, the top-individual by surprisal collected $20$ discoveries out of $68$ eating places where he purchased. His expected number of discoveries under the null model was $2.56$, meaning that he collected roughly eight times more discoveries than expected by chance. The probability he achieved $20$ discoveries under the null model is $4.7\times 10^{-13}$, resulting in a large surprisal value ($s=28.4$).
 
 \section{The bootstrap procedure}
 \label{sec:bootstrap}
 
 Even in a random sampling process, some individuals might still achieve a large value of surprisal due to statistical fluctuations. To ascertain that the largest observed values of surprisal cannot be explained by chance, we perform a bootstrap analysis~\citep{medo2016identification}. In each realization of the bootstrap procedure, for each individual, we extract its number $d_i$ of discoveries from the hypergeometric distribution~\eqref{hypergeo}, and compute the surprisal value, $s_i$, associated with the extracted number of discoveries. For each bootstrap realization, we rank the individuals according to their surprisal, obtaining the Zipf's plot $s(r)$ for that realization. We then average the Zipf plot obtained over different realizations, and compare the resulting Zipf plot with the Zipf plot corresponding to the empirical surprisal values. The results show that for the highest ranking positions, the empirical surprisal values are significantly larger than the bootstrapped ones  (Figs.~\ref{fig:fig1}E, S1--S2).
 
 \section{Statistical analysis}
 \label{sec:statistical}
 
 The statistical significance of the differences between group and population median displayed in Fig.~\ref{fig:fig4} have been obtained through the Binomial test and Mood’s median test~\cite{mood1950introduction} for the female ratio and all the other traits, respectively.
The binomial test allows us to test the null hypothesis that males and females are equally likely to occur in the detected group of individuals. The Mood's median test allows us to test whether the group and population median were extracted from two populations with the same median.

\subsection*{Materials and Data Availability}
Source code is available on request to the authors. For contractual and privacy reasons, the raw data is not available. Upon request, the authors can provide appropriate documentation for replication, and they might provide samples of the processed data.

\subsection*{Ethics declaration}
The Human Subjects Committee of the Faculty of Economics, Business Administration and Information
Technology at the University of Zurich has authorized this research on 29 March 2018. In particular, it has reviewed the information regarding the procedures and protocols in our research, and confirmed that they comply with
all applicable regulations.

\subsection*{Acknowledgements}

This work has been supported by the Science Strength Promotion Program of UESTC and by the URPP Social Networks at the University of Zurich. MSM and CJT acknowledge financial support from the Swiss National Science Foundation (Grant No. 200021-182659). MSM acknowledges financial support from the UESTC professor research start-up (Grant No. ZYGX2018KYQD215).

\clearpage
\onecolumngrid

\subsection*{References}

\let\oldbibliography\thebibliography
\renewcommand{\thebibliography}[1]{\oldbibliography{#1}
\setlength{\itemsep}{6pt}}

\bibliographystyle{apa}
\bibliography{bibliography-pnas}

\clearpage

\renewcommand{\thefigure}{A\arabic{figure}}
\renewcommand{\thetable}{A\arabic{table}} 
\setcounter{figure}{0}

\section*{Online Appendix A. Data filtering and networks construction}

Before describing the data, we point out that all the data analyzed in the article are anonymized. The subjects of the analysis (individuals and stores) are represented by meaningless hashes in the dataset. All individuals are non-identifiable, meaning that
 there is no way to reconstruct the individuals' real identities; all stores are nameless, there is no way to reconstruct the stores' real name; all transactions are innominate.
 For the sake of better readability, when referring to the analyzed datasets and the individuals in the following, we shall omit the labels "anonymized" and "non-identifiable".
 Nevertheless, whenever we will refer to an individual, it should be interpreted as a non-identifiable individual whose real identity is impossible to identify. Whenever we will refer to a group of "detected" individuals, it should be interpreted as a group of individuals whose real identities cannot be identified.

\begin{table}[t]
\begin{center}
\begin{tabular}{ |c|c|l|}
\hline
  Category & MCC & Description  \\
  \hline 
  Eating places & $5812$  & Restaurants or eating places    \\
                & $5813$  & Drinking Places (Alcoholic Beverages), Bars, Taverns, Cocktail lounges, Nightclubs and Discotheques          \\
                & $5814$  & Fast Food Restaurants           \\
  \hline
  Food stores & $5411$ & Grocery stores, Supermarkets \\
            & $5422$ & Freezer and Locker Meat Provisioners \\
            & $5441$ & Candy, Confectionery, Nut stores \\
            & $5451$ & Dairy Products stores \\
            & $5462$ & Bakeries \\
            & $5499$ & Misc. Food stores – Convenience stores and Specialty Markets \\
\hline
  Clothing stores & $5611$ & Men’s and Boy’s Clothing and Accessories stores\\
            & $5621$ & Women’s Ready-to-Wear stores     \\
            & $5631$ & Women’s Accessory and Specialty stores     \\
            & $5641$ & Children’s and Infant’s Wear stores     \\
            & $5651$ & Family Clothing stores    \\
            & $5655$ & Sports Apparel, Riding Apparel stores  \\
            & $5661$ & Shoe stores     \\
            & $5681$ & Furriers and Fur stores    \\
            & $5691$ & Men’s and Women’s Clothing stores     \\
            & $5697$ & Tailors, Seamstress, Mending, and Alterations     \\
            & $5698$ & Wig and Toupee stores      \\
            & $5699$ & Miscellaneous Apparel and Accessory stores     \\
\hline
\end{tabular}
\end{center}
\caption{Definition of the three store categories analyzed in this paper in terms of the Merchant Category Codes (MCCs) of the included stores.}
\label{tab:mcc}
\end{table}
 
\subsection*{Credit Card Record (CDR)}

We analyzed a Credit Card Record (CCR) from a large bank in an emerging country collected over a three-year period from June 2015 to June 2018. 
We filtered out stores with less than ten customers throughout the whole data time span.
We consider three store categories: eating places, food stores, and clothing stores. The three categories have been selected according to the Merchant Category Codes (MCCs) that are available in the data, according to the classification scheme reported in Table~\ref{tab:mcc}. 
We study separately three temporal bipartite networks where individuals are connected to the stores they purchased in. The time-stamp of each link is determined by the time-stamp of the first purchase by the individual. 

We split the three-year CCR into three non-overlapping periods, as explained below. The time periods reported below refer to the ones that were used for the analysis in the main text; in Online Appendix E, we tested the robustness of our results with respect to other choices for the data partitioning.
\begin{itemize}
    \item \textbf{Pre-filtering period (June 2015 -- November 2015)}. We used this period to assess whether a store that appears in the training period is a new store or a previously-existing one.
    If a store found in the training period is also found in the pre-filtering period, it is a pre-existing one and does not contribute to the customers' number of discoveries, whereas it still contributes to their number of visited stores. 
    \item \textbf{Training period (December 2015 -- May 2017)}. We analyze separately the three categories of stores reported in Table~S1. A potential issue is that the discoverer detection procedure requires us to estimate the success of the stores, which might be unreliable for stores that received their first transaction near the end of the training period. For this reason, we filtered out from the analysis the stores introduced less than two months before the end of the training period. The total number of relevant stores  is denoted as $S$. The time $t_{i\alpha}$ of each link $(i,\alpha)$ is determined by the first visit of individual $i$ to store $\alpha$. We denote as $k_i$ and $v_\alpha$ the number of stores visited by individual $i$ within the training period (excluding the last two months of this period) and the number of first-time customers of store $\alpha$, respectively. 
    We denote as $d_i$ the number of discoveries collected by individual $i$ within the training period (excluding the last two months of this period).
    The reason for excluding the last two months when measuring $k_i$ and $d_i$ is that the estimation of success might be unreliable for the shops that appeared for the first time near the end of the training period.
    \item \textbf{Validation period (June 2017 -- May 2018)}. The transactions from June 2017 to May 2018 are used as validation period to assess the out-of-sample predictive signal for different groups of detected individuals. We focus on the stores that were opened within this period, and assess the relation between the presence/absence of different groups of individuals among the earliest customers and the future success of the store (see Online Appendix D for the details). Again, a potential issue is that the prediction evaluation procedure requires us to estimate the future success of the stores, which might be unreliable for stores opened near to the end of the validation period. For this reason, we filtered out from the analysis the stores introduced less than two months before the end of the validation period. We denote by $V$ the resulting number of relevant stores. 
\end{itemize}

\subsection*{Call Data Record (CDR) and its relation with the CCR}

We analyzed a Call Data Record (CDR) from a large mobile phone operator from the same country where the CCR was recorded. The CDR used in this study covers a one-year period from January 2016 to December 2016. Importantly, this period overlaps with the CCR's time span, and it is possible to partially match the individuals in the CDR with the individuals in the CCR (see below). For each telco customer, the CDR contains all the calls she made to or received from both other telco customers and non-customers.
This implies that for the telco customers, we can observe their complete mobile-phone communication activity, whereas for the non-customers, we can only see their communications with the telco customers. Besides, each individual may be telco customer only for specific months of the year, but not throughout the whole year.

In our work, we use the CDR to construct $12$ snapshots of the social network. For each month, we only include the telco customers in the network.
When computing the time-averaged centrality metrics of each individual in the social network, we only include the months when the target individual was a telco customer. The rationale is that the individuals' centrality is largely underestimated in the months when they are not telco customers, because their calls from/to non-telco customers are not included. We refer to Online Appendix C for a description of how the time-averaged centrality metrics were computed.


\section*{Online Appendix B. Individual-level traits extracted from the CCR}
  
We describe here how we extracted the individual-level traits of interest from the available CCR.

\begin{itemize}
\item \textbf{Surprisal (Discoverers).}

The individuals' surprisal is defined by Eq. (2) in the main text. We refer to the main text for the details of its computation. The \textbf{discoverers} of a given category of stores are the top-$\tau\%$ individuals by the surprisal obtained by analyzing the respective category of stores.

\item \textbf{Number of visited stores (Store explorers).}

For each of the three categories of stores considered in our work, we count the number of visited stores per individual, $k$, within the training period. 
The \textbf{store explorers} of a given category of stores are the top-$\tau\%$ individuals by number of visited stores within the training period.

\item \textbf{Time-averaged total expenditure (High-expenditure individuals).}

For each bank customer who made at least one purchase within the training period, we extract his/her total expenditures from each month between January and December 2016, and we average over this $12$-month period. The \textbf{high-expenditure individuals}  are the top-$\tau\%$ individuals by time-averaged total expenditure.
\end{itemize}
 
\section*{Online Appendix C. Individual-level traits extracted from the CDR}

We describe here how we extracted the individual-level traits of interest from the available CDR. The extracted traits are used both to detect the top-individuals used to make predictions (as detailed below), and to characterize the groups of top-individuals in Fig.~3 of the main text.

\subsection*{Centrality and social hubs}

We introduce here appropriate time averages of three different centrality metrics: degree, collective influence, social diversity.
\begin{itemize}
\item \textbf{Time-averaged number of contacts (Social hubs by degree).}

The number of contacts (or degree) of the individuals in the social network is probably the simplest metric to quantify individuals' centrality. Individuals with a large number of contacts -- \textit{social hubs} -- have been first used for success prediction by Goldenberg \textit{et al.}~\citep{goldenberg2009role}.
To detect the social hubs~\citep{goldenberg2009role}, we measure the number of contacts per individual within each month, $k_i(t)$. We denote by $\mathcal{T}(t)$ the set of telco customers in month $t$. An individual may be telco customers only for some months within the CDR timespan. Motivated by the lines of reasoning in Online Appendix A, we define the time-averaged number of contacts for individual $i$ as
\begin{equation}
    \overline{k_i}=\frac{\sum_t \delta(i\in\mathcal{T}(t))\,k_i(t)}{\sum_t \delta(i\in\mathcal{T}(t))}.
\end{equation}
From the definition, it follows that $\overline{k_i}>0$ only for those individuals who are telco customers for at least one month. Only the months when an individual is a telco customer are included in the average.
The \textit{social hubs by degree} are the top-$\tau\%$ individuals by $\overline{k_i}$, among the individuals who are found in the CCR and made at least one purchase in stores of the analyzed category within the training period.

We note that if an individual is found in the CCR but she is not among the telco customers, she obtains $\overline{k}_i=0$ and cannot be detected as a social hub. This is a consequence of the incompleteness of our data: given a set of individuals who make transactions in the CDR, we do not know the communication activity for all of them and, as a result, we can only detect the top-individuals by centrality among those that are also telco customers. Similar remarks apply for all other individuals' traits extracted from the CDR.

\item \textbf{Time-averaged Collective Influence (Social hubs by Collective Influence, CI).}

While the degree is the simplest centrality metric in networks~\citep{lu2016vital}, it neglects higher-order network effects that are potentially informative about the position of the nodes. As a more sophisticated metric of network centrality, we rely on the \textit{collective influence} metric introduced by Morone and Makse~\citep{morone2015influence}. The metric detects the minimal set of nodes that, once removed from the network, disrupt the network's giant component. The detection of these nodes is typically referred to as the structural influence maximization problem~\citep{lu2016vital}. Morone and Makse~\citep{morone2015influence} solved analytically the problem through the theory of optimal percolation on graphs, and showed that the collective influence metric provides a reliable approximation to the solution of the problem and, at the same time, can be computed rapidly on large datasets.
 By considering the network of telco customers in month $t$, the collective influence $CI_i(t)$ of a telco customer $i$ in month $t$ is given by~\citep{morone2015influence}
     \begin{equation}
        CI^{(l)}_i(t)=(k_i(t) - 1)\sum_{j \in \partial B_{l}(i)}(k_j(t) - 1),
    \end{equation}
   where $B_{l}(i)$ denotes the ball of radius $l$ centered in $i$, and $\partial B_{l}(i)$ denotes its frontier~\citep{morone2015influence}. Here, we set $l = 1$.
As we did for the number of contacts, we define the time-averaged collective influence of individual $i$ as
    \begin{equation}
    \overline{CI_i^{(1)}}=\frac{\sum_t \delta(i\in\mathcal{T}(t))\,CI^{(1)}_i(t)}{\sum_t \delta(i\in\mathcal{T}(t))}.
\end{equation}
The \textit{social hubs by collective influence} are the top-$\tau\%$ individuals by $\overline{CI_i^{(1)}}$, among the individuals who are also found in the CCR and made at least one purchase in stores of the analyzed category within the training period.

\item \textbf{Time-averaged social diversity (Social hubs by social diversity)}

We consider an alternative metric of social importance that has brought insights into regional socioeconomic development~\citep{eagle2010network}.
The social diversity metric quantifies whether a given individual tends to communicate repeatedly with a restricted set of contacts, or whether she contacts a diverse set of people.
It is defined as the entropy~\citep{eagle2010network}
\begin{equation}
S_i^S(t)=    -\sum_j \frac{(w_{ij}(t)/w_i(t))\,\log(w_{ij}(t)/w_i(t))}{\log(w_i(t))}
\end{equation}
where $w_i$ represents $i$'s total number of interactions within month $t$, $w_{ij}(t)$ the total number of interactions between $i$ and $j$ within month $t$.
 If a person $i$ has only one contact $j$ over one month $t$, then $w_{ij}(t)=\delta_{ij}$ and, as a consequence, $S_i^S(t)=0$.
As we did for the previous centralities, we define the time-averaged social diversity of individual $i$ as
    \begin{equation}
    \overline{S_i^{S}}=\frac{\sum_t \delta(i\in\mathcal{T}(t))\,S^{S}_i(t)}{\sum_t \delta(i\in\mathcal{T}(t))}.
\end{equation}
 The \textit{social hubs by social diversity} are the top-$\tau\%$ individuals by $\overline{S^S}$, among the individuals who are also found in the CCR and made at least one purchase in stores of the analyzed category within the training period. 
\end{itemize}

\subsection*{Mobility-related traits and explorers}

\begin{itemize}
\item \textbf{Time-averaged mobility diversity}

The mobility diversity metric has been introduced to characterize the predictability of human mobility~\citep{song2010limits}.
The idea is to understand whether a given individual uses a restricted number of antennas or a diverse set of antennas. The latter means that the individual has been in a diverse set of locations, which is a manifestation of high mobility. The mobility diversity of individual $i$ is given by the following entropy~\citep{song2010limits}:
\begin{equation}
  S_i^{M}(t)=  -\sum \frac{ (c_{ia}(t)/c_i(t))\,\log(c_{ia}(t)/c_i(t))}{\log(c_i(t))}
\end{equation}
where $c_{ia}(t)$ denotes the number of calls made/received by $i$ through antenna $a$ within month $t$, and $c_i$ denotes the total number of calls made/received by individual $i$ within month $t$.
If $i$ uses only one antenna within month $t$, $c_{ia}(t)=c_i(t)$ which results in $S_i^{M}(t)=0$. On the other hand, individuals who made/received calls from many different antenna are characterized by large values of $S_i^{M}(t)$.
As we did for the previous CDR-extracted traits, we define the time-averaged mobility diversity of individual $i$ as
    \begin{equation}
    \overline{S_i^{M}}=\frac{\sum_t \delta(i\in\mathcal{T}(t))\,S^{M}_i(t)}{\sum_t \delta(i\in\mathcal{T}(t))}.
\end{equation}
  The \textit{explorers by mobility diversity} are the top-$\tau\%$ individuals by $\overline{S^{M}}$, among the individuals who are also found in the CCR and made at least one purchase in stores of the analyzed category  within the training period. 
  
\item \textbf{Time-averaged radius of gyration}

The radius of gyration can be interpreted as the characteristic distance traveled by a given individual~\citep{gonzalez2008understanding}. This metric has been used to distinguish explorative individuals from "returner" individuals who tend to only visit a small number of locations~\citep{pappalardo2015returners}.
Individual $i$'s total radius of gyration $r_{g}$ is defined as~\citep{gonzalez2008understanding}
\begin{equation}
r_{i} = \sqrt{ \frac{1}{N_i}\sum_{l \in \mathcal{L}_i}{n_{il} \,\Bigr(\mathbf{r}(l) -\mathbf{r}^{CM}_{i}\Bigl)^2}},
\end{equation}
where $n_{il}$ denotes the number of times individual $i$ uses antenna $l$ within month $t$, $\mathcal{L}_i$ denotes the set of antennas that individual $i$ visited, $N_i = \sum_{l \in \mathcal{L}} n_i$, $\mathbf{r}(l)$ is a two-dimensional vector describing the geographic coordinates of location $l$, $\mathbf{r}^{CM}_{i}$ represents individual $i$'s center of mass.
As we did for the previous CDR-extracted traits, we define the time-averaged radius of gyration of individual $i$ as
    \begin{equation}
    \overline{r_i}=\frac{\sum_t \delta(i\in\mathcal{T}(t))\,r_i(t)}{\sum_t \delta(i\in\mathcal{T}(t))}.
\end{equation}
  The \textit{explorers by radius of gyration} are the top-$\tau\%$ individuals by $\overline{r}$, among the individuals who are also found in the CCR and made at least one purchase in stores of the analyzed category  within the training period.

\end{itemize}

\section*{Online Appendix D. Success prediction: formulation of the problem}

We provide here all details on the formulation of the predictive problem, the Naive Bayes Classifiers used in the paper, and the prediction evaluation metrics.

\subsection*{Formulation of the predictive problem}

In line with the literature on the prediction of popularity in online systems~\citep{martin2016exploring,shulman2016predictability,hofman2017prediction}, we aim to address the following question: Can we use the behavioral and socioeconomic traits of stores' early customers to predict the eventual popularity of the store? 
In other words, we \textit{peek} into the stores' early activity data, and we aim to use this information to predict the stores' eventual success.
Formulating the related predictive problem requires two design choices: (1) How much peeking into early activity in a store is allowed, and (2) which metric of store success we aim to predict~\citep{shulman2016predictability,hofman2017prediction}.

As for (1), ideally, we would like to look at a period of early activity that is short enough that the eventual success of the store is not evident.
Besides, we would like to exclude from the predictive problem stores that only received few customers and, for this reason, are unlikely to become successful in the future.
Motivated by these considerations, we peek into the stores' earliest $w$ first-time customer, where we vary $w$ from $5$ to $30$; when we consider the earliest $w$ first-time customer, only stores that received at least $w$ first-time customer are included in the analysis.

As for (2), a broad range of predictive problems emerges depending on the dimension of store success that we aim to predict. In this work, we focus on the stores' popularity, $v$, defined by their total number of first-time customers (i.e., the total number of people who purchase in the store at least once). However, we cannot use directly $v$ to define the group of successful stores, as $v$ is strongly influenced by store age and MCC (see Online Appendix B). One faces an analogous issue when measuring the impact of scientific papers through citation-based indicators (which often results in metrics of impact that are biased by paper age and scientific field~\citep{liao2017ranking}), or when attempting to compare the performance of different innovation diffusion processes that started at different points in time~\citep{muller2019effect}.
To factor out these confounding effects, we define the group $\mathcal{P}$ of popular stores as the group of stores that are ranked among the top-$s\%$ stores by cumulative number of first-time customers, \textit{among stores with the same MCC that received their first transaction in the same month}. The cumulative number of first-time customers is measured throughout the complete validation period. Assessing the popularity of each store only in relation to stores of the same MCC and age directly removes the bias of $v$ without the need to know the distribution of $v$.

Putting together (1) and (2), we formulate a binary classification problem where we observe the earliest $w$ first-time customers of a store within the validation period (excluding the last two months), and we aim to predict whether the store will end up in the group $\mathcal{P}$ of the popular stores. Importantly, to validate the predictions, we only consider stores that received their first purchase during the validation period (excluding the last two months), whereas the eight different classes of top-individuals defined in the main text and Online Appendix B are detected within the training period.

\subsection*{Naive Bayes Classifiers (NBCs)}

To quantify the predictive performance of different groups of individuals, we consider \textbf{Naive Bayes Classifiers} (NBCs)~\citep{sarigol2014predicting}. For each group $\mathcal{I}$ of relevant individuals (e.g., $\mathcal{I}$ can represent the set of discoverers), we consider the simplest possible classification rule: \textit{a store is classified as successful if at least one individual $i\in\mathcal{I}$ was among the earliest $w$ customers, as non-successful otherwise}. Such a simple classifier allows us to compare the predictive power of the eight groups of relevant individuals considered here. From a machine-learning standpoint, this classifier requires no model training: it only takes as input the list of individuals detected within the training period. It is useful to introduce a random classifier that provides us, for each prediction evaluation metric, with a baseline performance metric. Such a random classifier is simply a random guess: each store is classified as successful with probability $0.5$, as non-successful with probability $0.5$.

Beyond this simple classifier, we also consider "two-dimensional" NBCs that result from the joint presence of individuals from two groups of relevant individuals.  For each pair $(\mathcal{I}_1,\mathcal{I}_2)$ of relevant individuals (e.g., $\mathcal{I}_1$ and $\mathcal{I}_2$ may represent the set of discoverers and social hubs, respectively), we implement the following classification rule: \textit{a store is classified as successful if and only if at least one individual from $\mathcal{I}_1$ and one individual from $\mathcal{I}_2$ were among the earliest $V$ first-time customer, as non-successful otherwise}. Compared to the $1$-dimensional NBCs, this rule is stricter, which results in a smaller number of stores classified as successful. As shown in Fig.~A6, this can result in precision improvements that range from marginal to large, depending on the pairs of considered individuals and store category.

\subsection*{Prediction evaluation metrics}

The performance in the classification task is measured in terms of the four elements of the confusion matrix (also referred to as contingency table)~\citep{powers2011evaluation,james2013introduction,chicco2017ten}: number of True Positives ($T^+$), False Positives ($F^+$), True Negatives ($T^-$), False Negatives ($F^-$).
We consider the following evaluation metrics:
\begin{itemize}
    \item \textbf{Precision} or \textbf{success rate}~\citep{powers2011evaluation}. Fraction of positive-classified stores that ended up in $\mathcal{P}$. In formulas, the precision $P$ is given by~\citep{powers2011evaluation}
    \begin{equation}
        P=\frac{T^+}{T^{+} +F^+}.
    \end{equation}
    Therefore, the ratio between the precision for the NBC of group $\mathcal{I}$ and the expected precision of the random classifier (or equivalently, the ratio between $\mathcal{I}$'s success rate and the baseline success rate) can be interpreted as the \textit{fold increase in success rate}\footnote{The terminology "fold change" is typically used in biology. In our study, it is particularly convenient because it describes effectively the fact that the stores with individuals from a given group of individuals among their early customers may exhibit increased odds of success.} (see main text). Groups of individuals with success-rate fold increase substantially larger than one can be interpreted as predictors of success.
    \item \textbf{Recall.} The recall, $R$, is commonly used in information retrieval and it is typically considered as a complementary metric to the precision. It is defined as the fraction of successful items that are labeled as positive~\citep{powers2011evaluation}
    \begin{equation}
    R=\frac{T^+}{T^+ + F^-}.
    \end{equation}
    In our study, the recall metric naturally favors groups of individuals who purchased in many different stores, because they are more likely to label a store as positive and, therefore, to label the successful ones as positive. However, the metric is still informative to have a full understanding of the predictions by the different groups of individuals: While in the main text we assessed whether stores that received an early purchase by a discoverer are more likely to be successful, the recall metric informs us about how many successful stores we can hope to detect by tracking the discoverers.
    \item \textbf{Positive likelihood ratio.} The positive likelihood ratio, $L$, is commonly used in medicine and diagnostic testing~\citep{mcgee2002simplifying}. For our problem, it is defined as the probability that a positive-classified store \emph{does} end up in $\mathcal{P}$ divided by the probability that a negative-classified store \emph{does} end up in $\mathcal{P}$. It can be expressed in terms of recall and specificity~\citep{powers2011evaluation}:
    \begin{equation}
        L=\frac{R}{1-S}.
    \end{equation}
    where $R=T^+/(T^+ + F^-)$ denotes the recall (or true positive rate) and $S=T^-/(T^- + F^+)$ denotes the specificity (or true negative rate).
    \item \textbf{Matthews' correlation coefficient ($r$)}. As the number of popular stores, $|\mathcal{P}|$, is substantially smaller than the total number of stores, traditional prediction-evaluation metrics like accuracy and the F1-score become ineffective to evaluate classifiers' predictive performance. 
    It has been shown numerically~\citep{boughorbel2017optimal,luque2019impact} that the Matthews' correlation coefficient is substantially less sensitive to data imbalance than the accuracy and the F1-score and, therefore, it is preferable~\citep{chicco2017ten}. The Matthews' correlation coefficient, $r$, is defined by the equation~\citep{matthews1975comparison}
    \begin{equation}
        r=\frac{T^+\,T^- - F^+\,F^-}{\sqrt{(T^+ + F^+)\,(T^+ + F^-)\,(T^- + F^+)\,(T^- + F^-)}}.
    \end{equation}
    Differently from the accuracy and F1 score, $r$ is robust with respect to variations of relative class size~\citep{boughorbel2017optimal}, which makes it suitable for our classification task that features imbalanced classes~\citep{chicco2017ten}.
\end{itemize}

\subsection*{Results}

The full results for the parameters adopted in the main text are reported in Fig.~A5, and the results for different parameter settings are reported in Figs.~A7--A13 and discussed in Online Appendix E.
Here, we briefly discuss the results in Fig.~A5.
The results for the success-rate fold increase were already shown in Fig.~3 in the main text and discussed there.
The positive likelihood ratio is in almost perfect agreement with the success-rate fold increase. The Matthews' correlation coefficient is also in good agreement with the success-rate fold increase, although some evident deviations can be observed: for example, store explorers and high-expenditure individuals emerge clearly as the second best-performing group of individuals for eating places and food stores, respectively. The results for recall significantly differ, which reflects the fact that metric rewards highly-active individuals who purchase in many different stores. The store explorers emerge as the best-performing group, followed by the discoverers and the high-expenditure individuals.

In the main text, we focused on the simplest possible predictive classifier based on the early purchases by different groups of individuals. We report here that precision improvements are possible through combinations of these features.
The two-dimensional classifiers introduced above tend indeed to outperform the one-dimensional classifier based on the discoverers in terms of success rate (Fig.~A6). For example, combining the discoverers with one of the three groups of explorers or high-expenditure individuals leads to a success-rate improvement for all three store categories. 
On the other hand, combining the discoverers with one of the three groups of social hubs leads to a success-rate improvement only for eating places and food stores, but not for clothing stores. This is unsurprising given the poor performance of the three groups of social hubs when considered independently (Fig.~3 in main text), and it suggests that combining pairs of groups of individuals can marginally improve the success rate, but only if both groups have an above-baseline success rate when considered alone.

\section*{Online Appendix E. Robustness of the predictions}

Our predictive analysis has five main parameters. 
Two parameters are specific to the discoverers detection method ($\Delta,z$), and three parameters are related to the predictive problem design ($s,\tau, \Delta t_{id}$). We provide below a description of the parameter variations that we performed, and an overview of the results. 
We begin by the two parameters of the discoverers detection method:
\begin{itemize}
    \item The duration of the time window over which discoveries can be collected (time window for discoveries), $\Delta$. By definition, a discovery is indeed only achieved when an individual purchases in a store no more than $\Delta$ days after the store received its first transaction. In the main text, we set $\Delta=90\,dd$. We subsequently tested $\Delta=60,120,\infty$ (see Fig. A7 for the results). Note that with $\Delta=\infty$, a successful store can be discovered throughout the whole training period, regardless of the time at which it receives its first transaction. The results (Fig.~A7) indicate that for $\Delta>90\,dd$, the predictive power is little sensitive to $\Delta$. On the other hand, a too narrow time window for discoveries ($\Delta=60\,dd$) leads to suboptimal performance.
    
    \item The percentage of stores that are considered as successful in the discoverer detection procedure, $z\%$. In the main text, we set $z\%=10\%$. We subsequently tested $z\%=5,20,30,100\%$ (see Fig.~A8 for the results). Interestingly, the discoverers detected with $z\%=100\%$ cannot be interpreted as discoverers of successful stores because with this value of $z\%$, all stores are candidates for discoveries. Therefore, the detected individuals are better interpreted as persistent early adopters.  The results are mixed: For eating and food stores, a more restrictive definition of success in the training period ($z\%=5\%$) can lead to a better out-of-sample performance in terms of precision, Matthews' correlation, and positive likelihood ratio. On the other hand, for clothing stores, a looser filter for successful stores ($z=20\%$) and even no filter at all ($z=100\%$) can achieve a better performance than that by the discoverers used in the main text ($z=10\%$). This is in qualitative agreement with the finding that explorative individuals perform well for clothing stores (see Fig.~3C).
\end{itemize}
We turn our attention to the three parameters related to the predictive problem design:
\begin{itemize}
    \item The percentage of stores that are considered as successful in the prediction evaluation, $s\%$. In the main text, we set $s\%=10\%$. We subsequently tested $s\%=5\%, 20\%$ (see Figs.~A9--A10 for the results). We make a conservative choice and always use the same parameters for the discoverer detection as in the main text ($z\%=10\%$); in principle, better performance might be achieved by tuning the value of $z\%$ to match $s\%$. There is still a positive predictive signal for the discoverers across the three categories. We note that the discoverers' signal weakens the most for clothing stores for a more selective success threshold ($s\%=5\%$, Fig.~A9); in this setting, the explorers by radius of gyration emerge as better predictors of success.
    \item The percentage of individuals selected as top-individuals by each metric, $\tau\%$. In the main text, we set $\tau\%=1\%$. We subsequently tested a more selective threshold, $\tau\%=0.5\%$ (see Fig. A11 for the results). The results are mostly consistent with those obtained in the main text.
    \item The relative duration of the training and validation period. By denoting with $\Delta t_{id}$ and $\Delta t_{val}$ the duration of the training and validation period, respectively, we started by performing all the analysis with $\Delta t_{id}=18\,mm$ (from Dec. 2015 to May 2017) and $\Delta t_{val}=12\,mm$ (from June 2017 to May 2018), as described in Online Appendix A. Subsequently, we repeated the analysis for a shorter training period ($\Delta t_{id}=16\,mm,\Delta t_{val}=14\,mm$, see Fig.~A12) and a longer one ($\Delta t_{id}=20\,mm,\Delta t_{val}=10\,mm$, see Fig.~A13). The results indicate that a longer validation period (and, correspondingly, a shorter training period) can lead to a stronger predictive power for the discoverers. In particular, in Fig.~A12 and for less than $15$ early customers included, the discoverers emerge as the top-performing individuals also for clothing stores. The discoverers' predictive power becomes weaker for a shorter validation period (Fig.~S13), yet the discoverers remain the top-performing individuals for eating places and food stores. 
\end{itemize}
 
To summarize, in our analysis, we started by performing the complete analysis with a predefined set of parameters ($\Delta=90\,dd,z\%=10\%,s\%=10\%,\tau\%=1\%,\Delta t_{id}=18\,mm$, results reported in Figs.~ of the main text and Fig.~A5), and subsequently assessed the robustness of our predictions against variations of the parameters (Figs.~A7--A13). The results are robust with respect to reasonable variations of the parameters, yet some observed variations are highly informative of the role of the various parameters. In particular, compared to the predictive results reported in the main text, factors that can improve the discoverers' predictive power are: calibrating the definition of success used in the discoverer detection (i.e., $z\%$); a longer validation period (and therefore, a larger number of stores used to validate the predictions); a more restrictive threshold ($\tau\%$) to select the top-individuals that are tracked to make the predictions.


\clearpage

\section*{Online Appendix F. Supplementary Figures}

\subsection*{Basic data properties}

    \begin{figure*}[h]
\centering
   \includegraphics[scale=0.35]{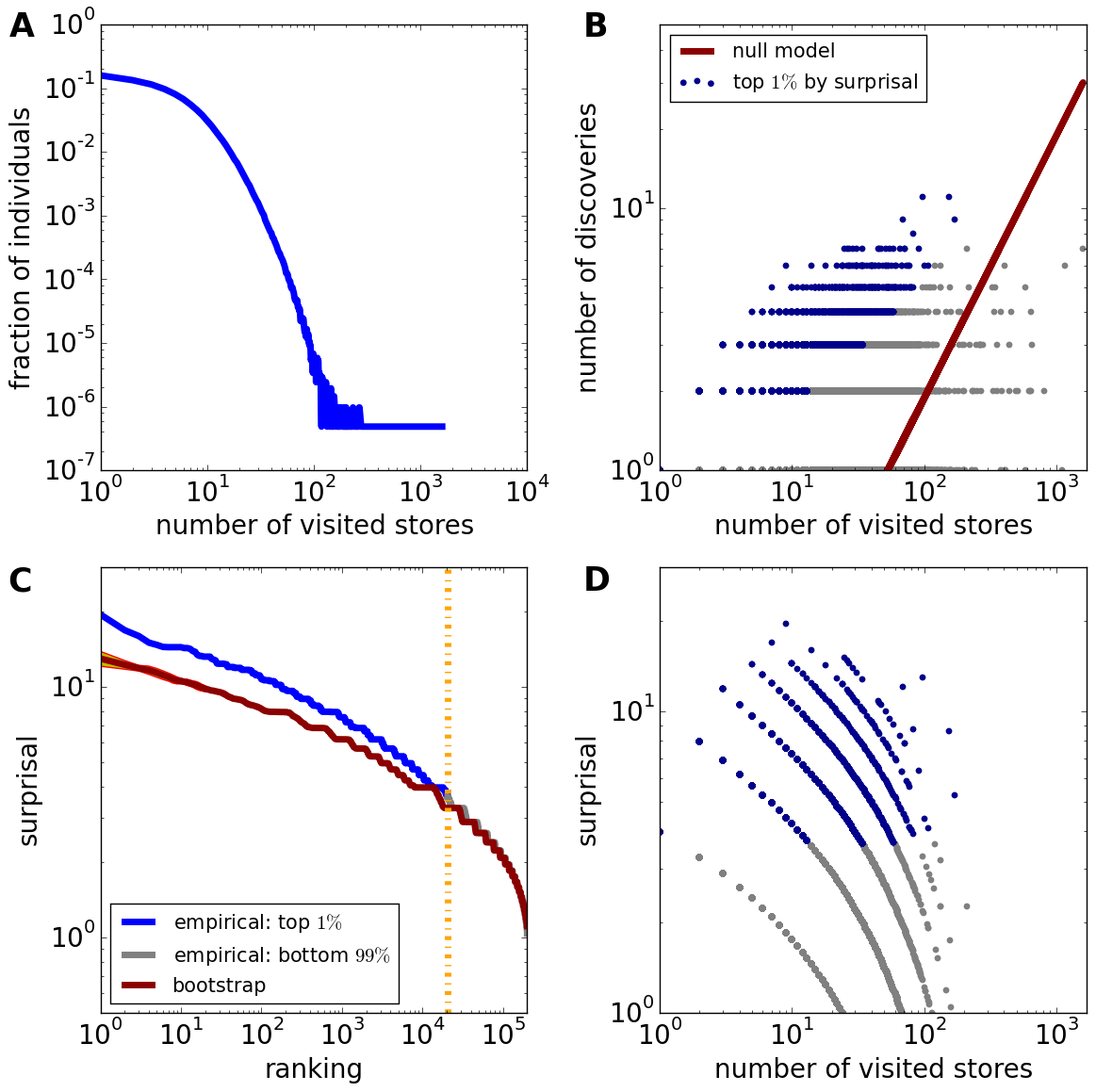}
   \caption{\textbf{Individuals' activity and discovery patterns for food stores}. (A) Distribution of the number of visited stores per individual. (B) The number of discoveries and number of visited stores per individual are positively correlated, yet the trend deviates from the expected one under the null model, and some individuals collect significantly more discoveries than expected from their activity alone. (C) Zipf plot of the individuals' surprisal for the empirical data and the null model. The discrepancy between the two curves is significant, meaning that the surprisal values of the top-$1\%$ individuals by empirical surprisal (i.e., those at the left of the vertical dashed line) are significantly above the surprisal values expected for their ranking position. (D) The individuals' surprisal is only weakly correlated with the number of visited stores.}
    \end{figure*}
    
        \begin{figure*}[h]
\centering
   \includegraphics[scale=0.35]{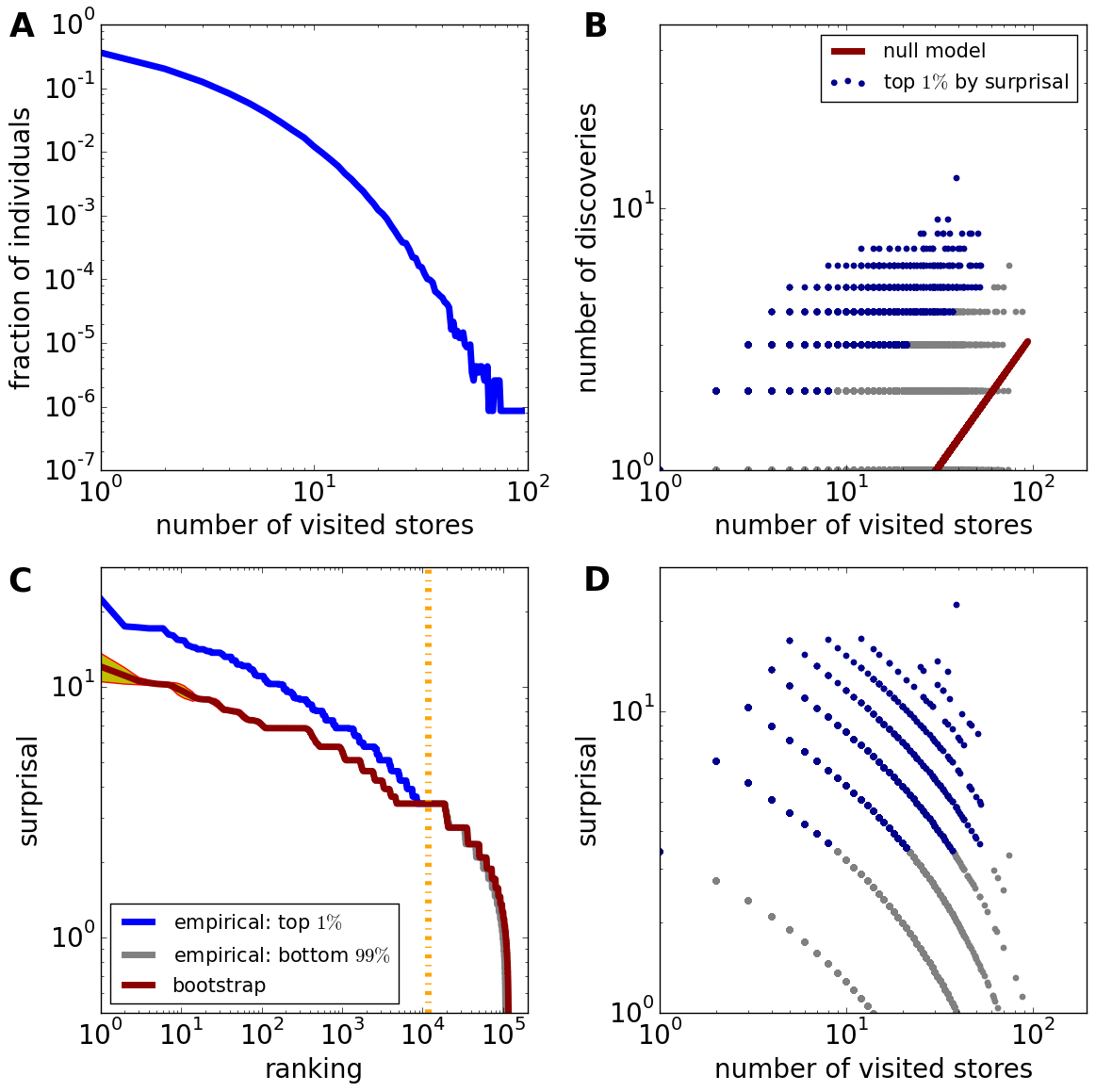}
   \caption{\textbf{Individuals' activity and discovery patterns for clothing stores}. (A) Distribution of the number of visited stores per individual. (B) The number of discoveries and number of visited stores per individual are positively correlated, yet the trend deviates from the expected one under the null model, and some individuals collect significantly more discoveries than expected from their activity alone. (C) Zipf plot of the individuals' surprisal for the empirical data and the null model. The discrepancy between the two curves is significant, meaning that the surprisal values of the top-$1\%$ individuals by empirical surprisal (i.e., those at the left of the vertical dashed line) are significantly above the surprisal values expected for their ranking position. (D) The individuals' surprisal is only weakly correlated with the number of visited stores.}
    \end{figure*}

  \begin{figure*}[h]
\centering
   \includegraphics[scale=0.25]{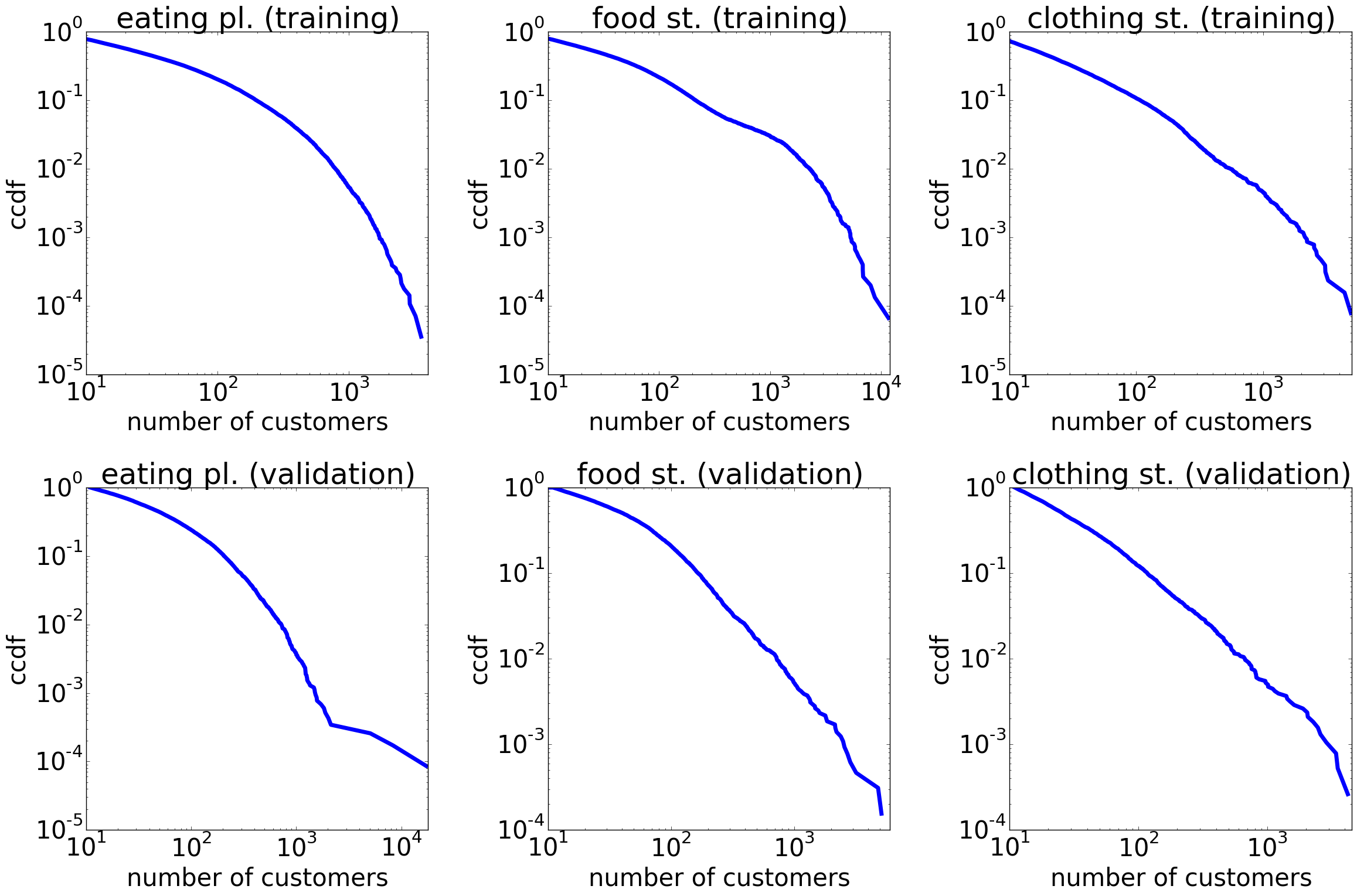}
   \caption{\textbf{Store popularity distributions.} Complementary Cumulative Distribution Function (CCDF) of the number of first-time customers per store, $c$, for stores that received their first transaction within the training period (top panels) and the validation period (bottom panels), respectively. As we filtered out stores with less than $10$ customers over the whole dataset, we only show the distribution in the range $c\geq 10$.}
    \end{figure*}
 
     \begin{figure*}[h]
\centering
   \includegraphics[scale=0.275]{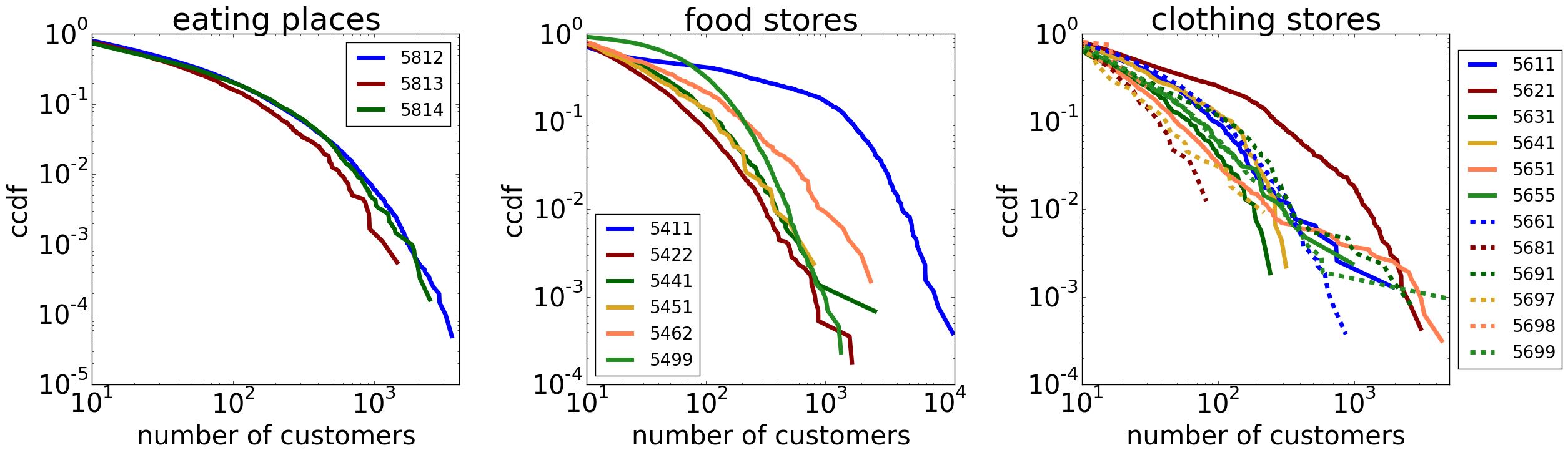}
   \caption{\textbf{Store popularity distributions for different Merchant Category Codes (MCCs).} Complementary Cumulative Distribution Function (CCDF) of the number of first-time customers per store, for stores that received their first transaction within the training period. We plot separately the distributions for stores that belong to different Merchant Category Codes (see Table S1). The results indicate that different MCCs exhibit different success distributions, which motivates our choice to compare only stores with the same MCC to detect the group of popular stores (see Material and Methods in the main text).}
    \end{figure*}
    
    
    \clearpage
    
    \subsection*{Prediction results: additional metrics}

\begin{figure*}[h]
\centering
   \includegraphics[scale=0.225]{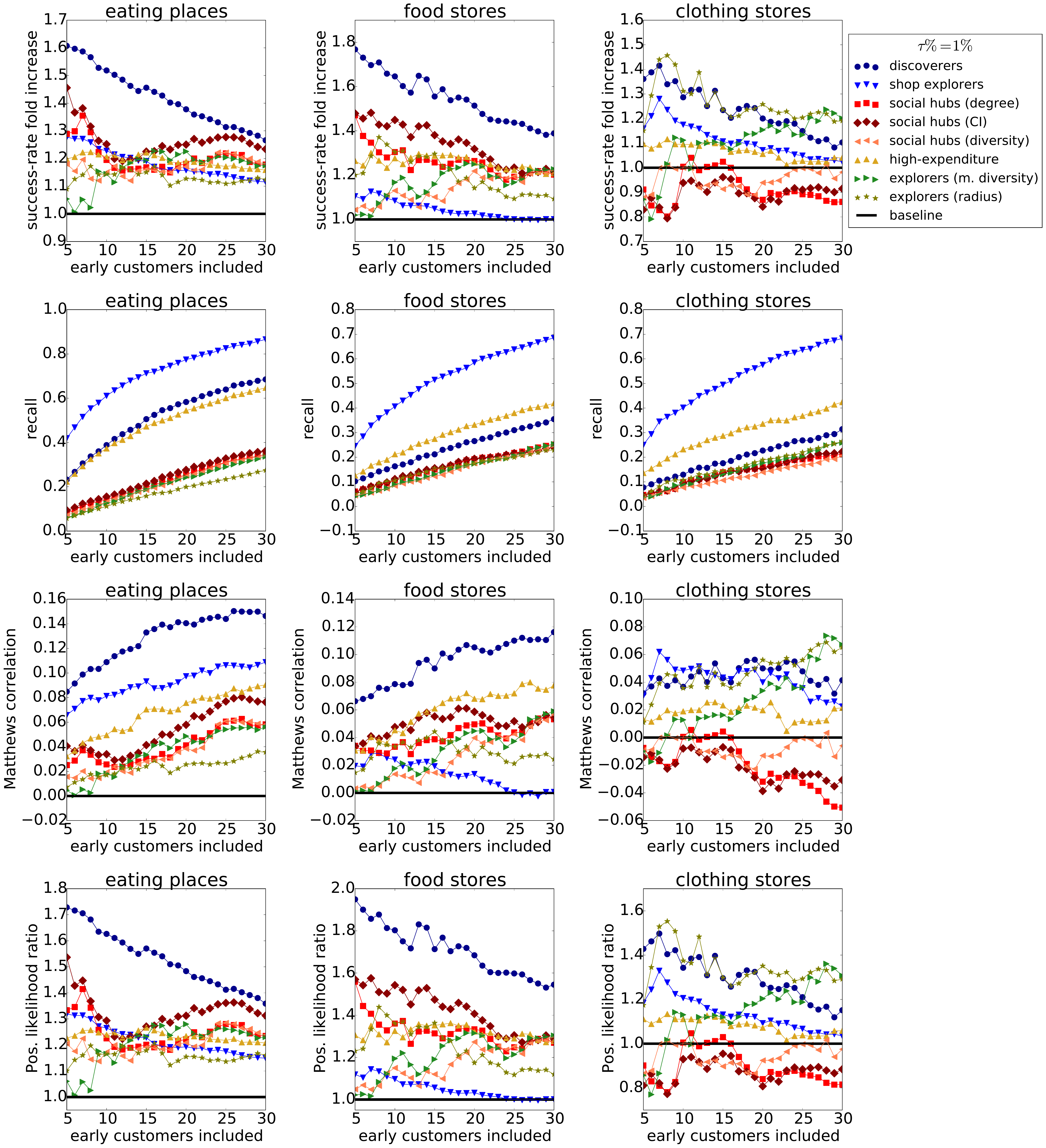}
   \caption{\textbf{Prediction results with the parameters adopted in the main text: $\Delta=90, z\%=10\%, s\%=10\%, \tau\%=1\%$}. }
    \end{figure*}

\clearpage 

\subsection*{Prediction results: Two-dimensional classifiers}

    \begin{figure*}[h]
\centering
   \includegraphics[scale=0.45]{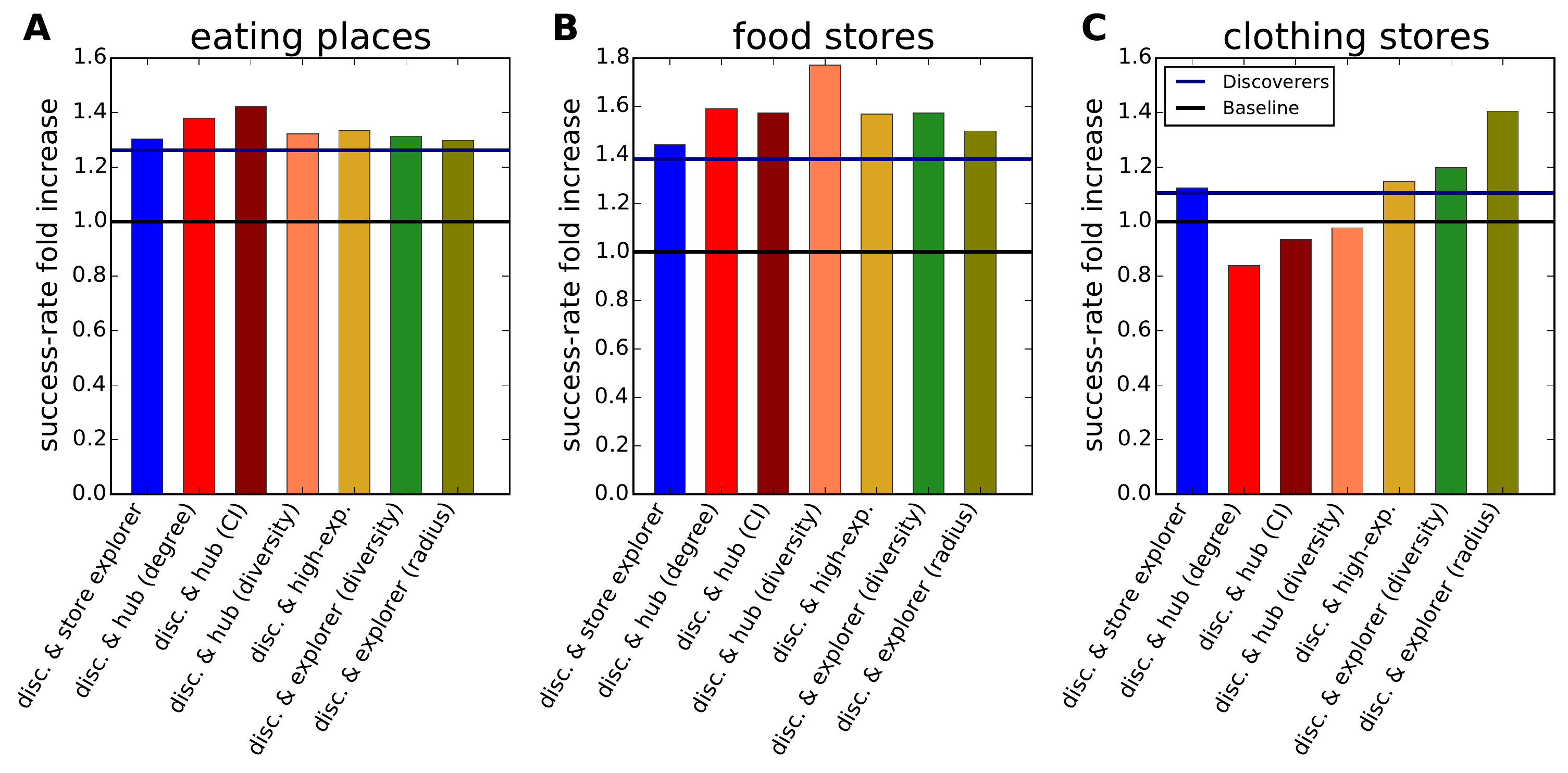}
   \caption{\textbf{Success predictions based on pairs of groups of key individuals.} Success-rate fold increase for stores that received an early purchase by individuals that belong to two different groups of top-individuals. We include the $30$ earliest customers here. Stores that received an early purchase by both a discoverer and an individual from another group of top-individuals can exhibit a substantially larger success rate than stores that received an only purchase by a discoverer. For example, stores that received an early purchase by both a discoverer and a high-expenditures individual exhibit a success-rate fold increase of $1.34,1.58,1.13$ for eating places, food stores, and clothing stores, respectively, representing improvements by $5.6\%, 14.9\%, 0.6\%$ with respect to the respective discoverers' success rates. }
    \end{figure*}

\clearpage

    \subsection*{Prediction results: varying $\Delta$}

\begin{figure*}[h]
\centering
   \includegraphics[scale=0.225]{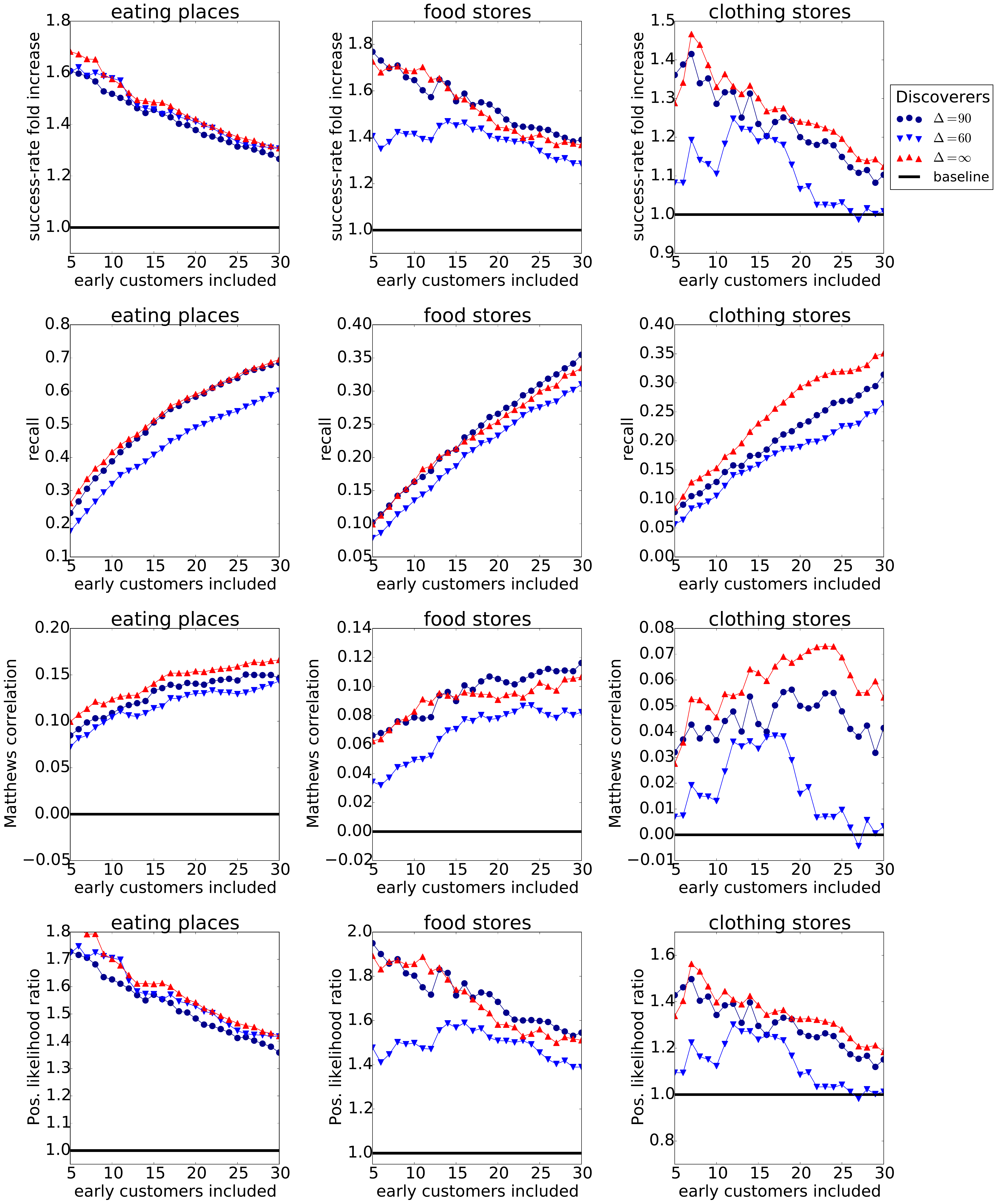}
   \caption{\textbf{Prediction results for different values of $\Delta$: $\Delta=60,90,\infty$} (with the other parameters being fixed as in the main text: $z\%=10\%, s\%=10\%, \tau\%=1\%$).}
    \end{figure*}

\clearpage

\subsection*{Prediction results: varying $z\%$}

\begin{figure*}[h]
\centering
   \includegraphics[scale=0.225]{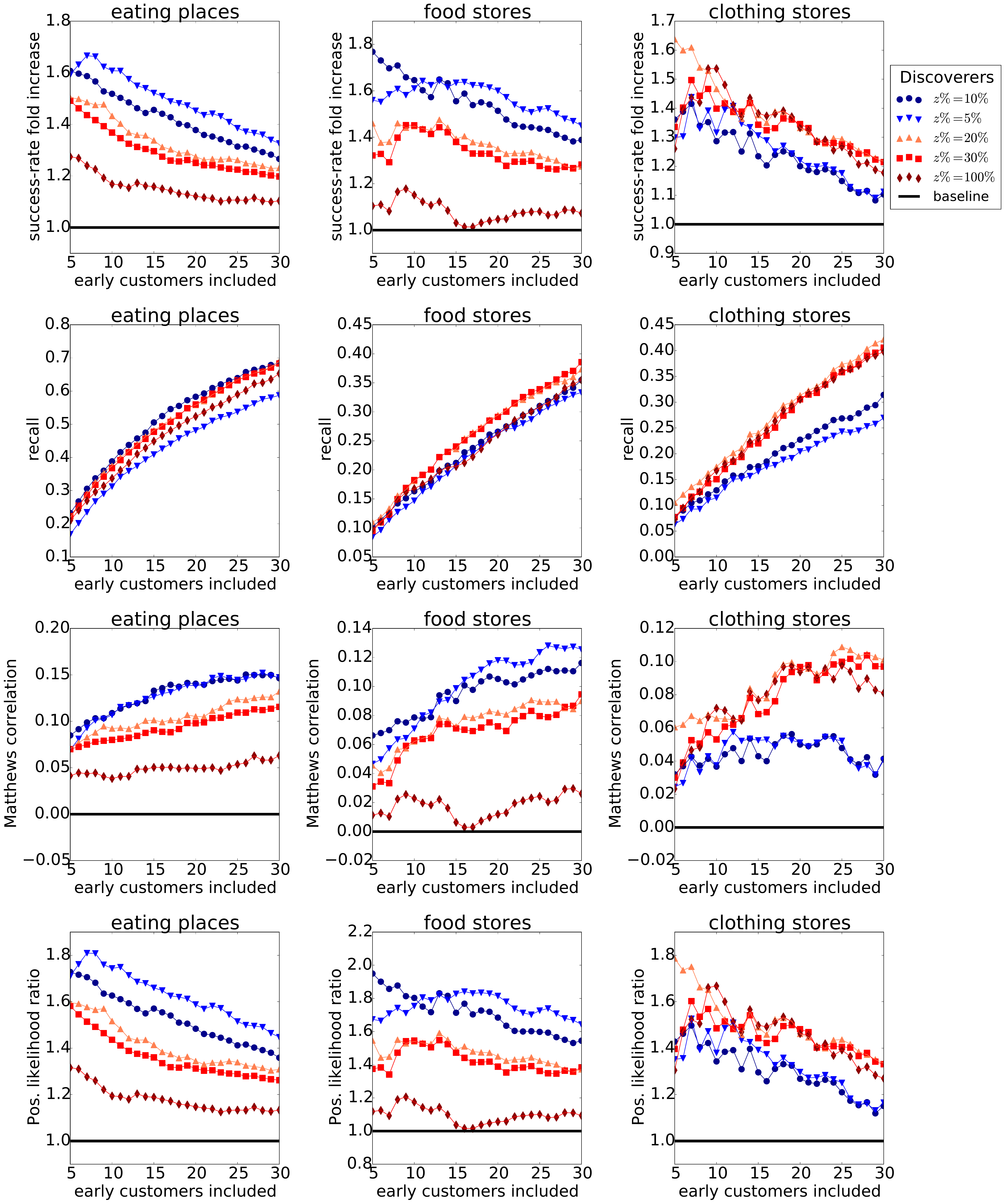}
   \caption{\textbf{Prediction results for different values of $z\%$: $z\%=10\%$ (main-text value), and $z\%=5,20,30,100\%$} (with the other parameters being fixed as in the main text: $\Delta=90\,\text{dd}, s\%=10\%, \tau\%=1\%$).}
    \end{figure*}

\clearpage 
\subsection*{Prediction results: varying $s\%$}

\begin{figure*}[h]
\centering
   \includegraphics[scale=0.25]{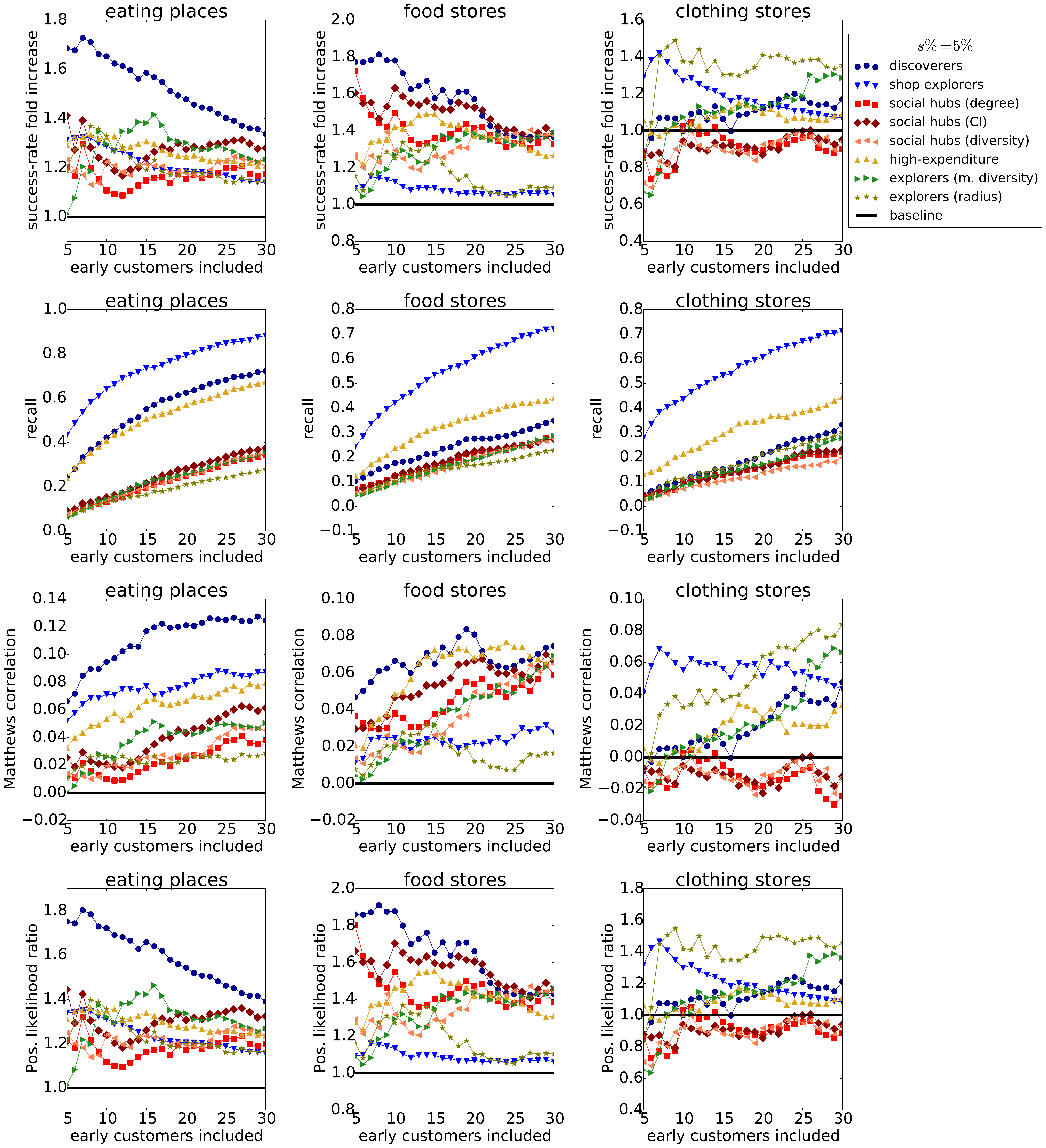}
   \caption{\textbf{Prediction results for a more selective value of $s\%$: $\Delta=90, z\%=10\%, s\%=5\%, \tau\%=0.5\%$}.}
    \end{figure*}
    
    \begin{figure*}[h]
\centering
   \includegraphics[scale=0.25]{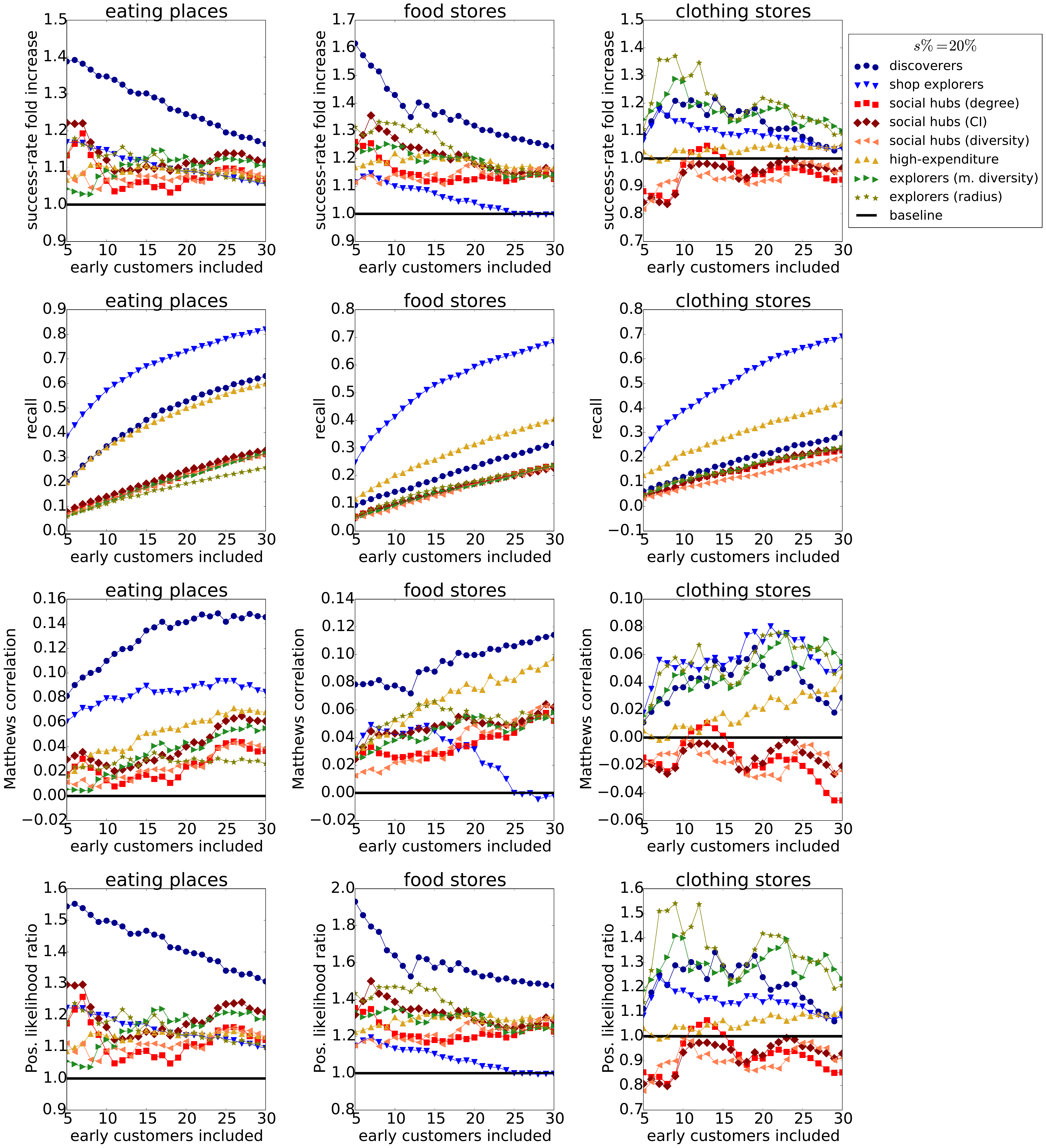}
   \caption{\textbf{Prediction results for a less selective value of $s\%$: $\Delta=90, z\%=10\%, s\%=20\%, \tau\%=0.5\%$}.}
    \end{figure*}

\clearpage
\subsection*{Prediction results: varying $\tau\%$}

\begin{figure*}[h]
\centering
   \includegraphics[scale=0.25]{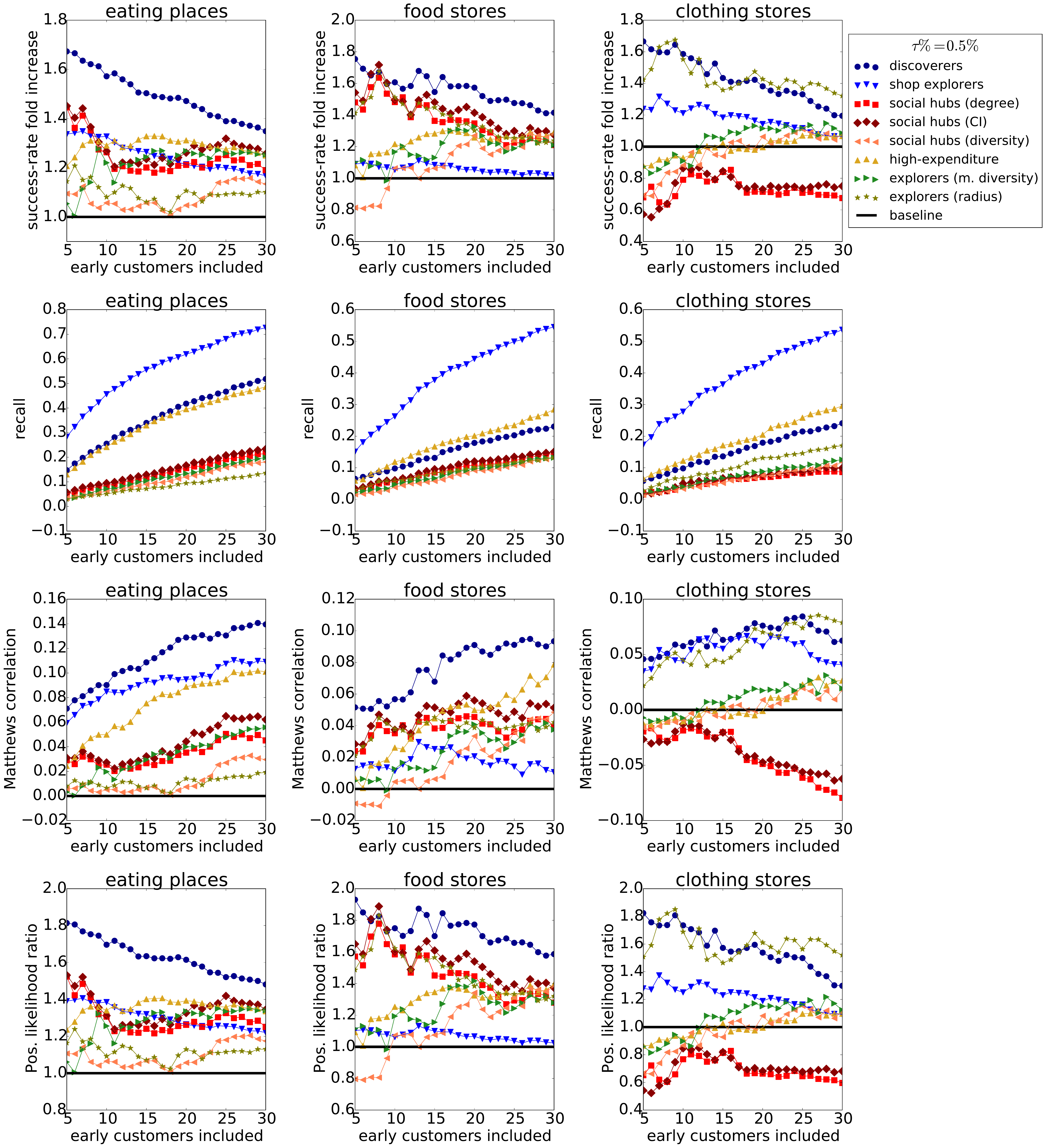}
   \caption{\textbf{Prediction results for a more selective value of $\tau\%$: $\Delta=90, z\%=10\%, s\%=10\%, \tau\%=0.5\%$}.}
    \end{figure*}
    
    \clearpage
    
\subsection*{Prediction results: varying the relative duration of the training and validation period}

\begin{figure*}[h]
\centering
   \includegraphics[scale=0.225]{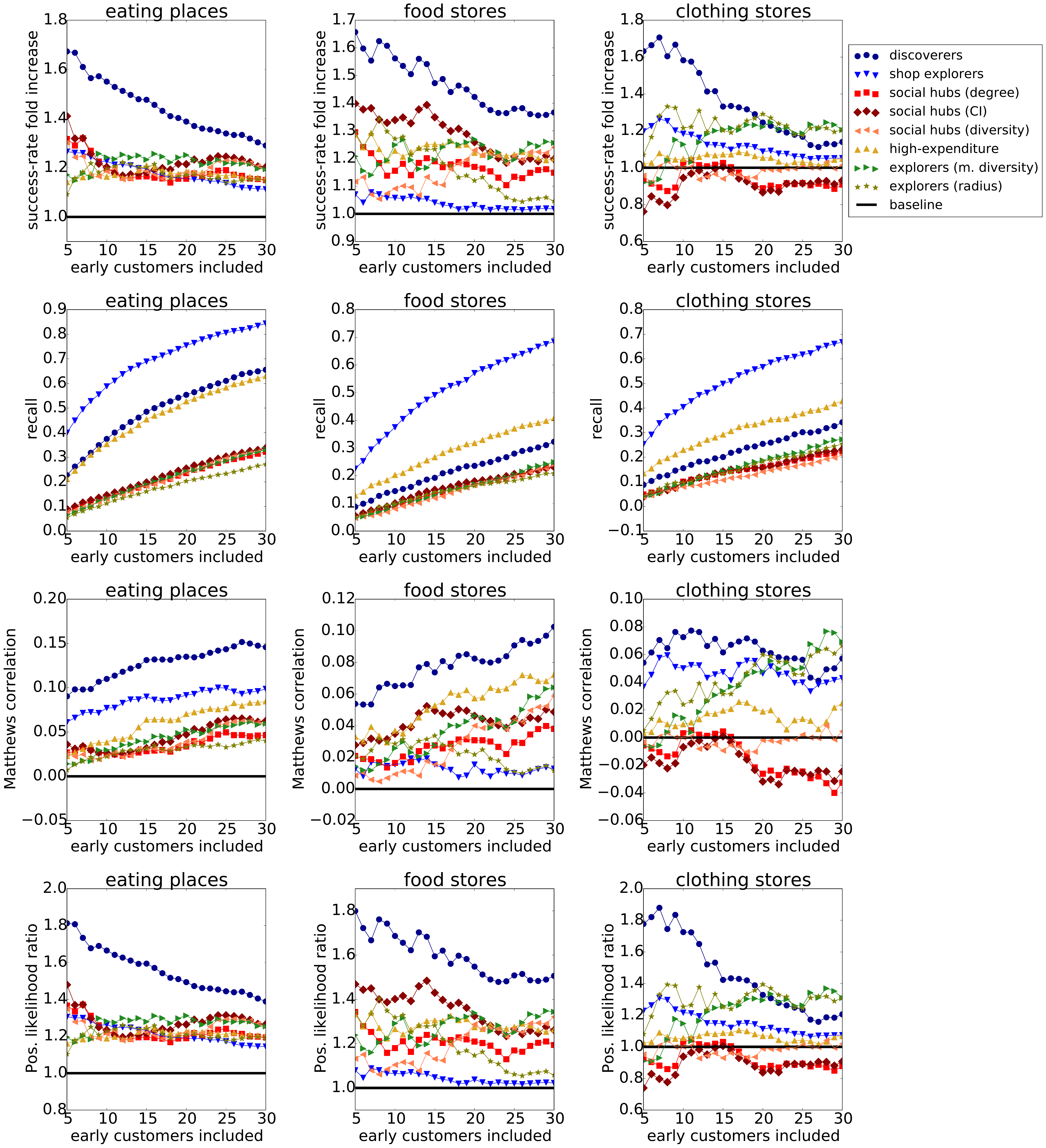}
   \caption{\textbf{Prediction results for a shorter training period (from Dec. 2015 to Mar. 2017, 16 months) and a longer validation period (from Apr. 2017 to May 2018, 14 months) compared to those used in the main text}. The other parameters of the analysis are the same as in the main text: $\Delta=90, z\%=10\%, s\%=10\%, \tau\%=1\%$.}
    \end{figure*}
    
    \begin{figure*}[h]
\centering
   \includegraphics[scale=0.225]{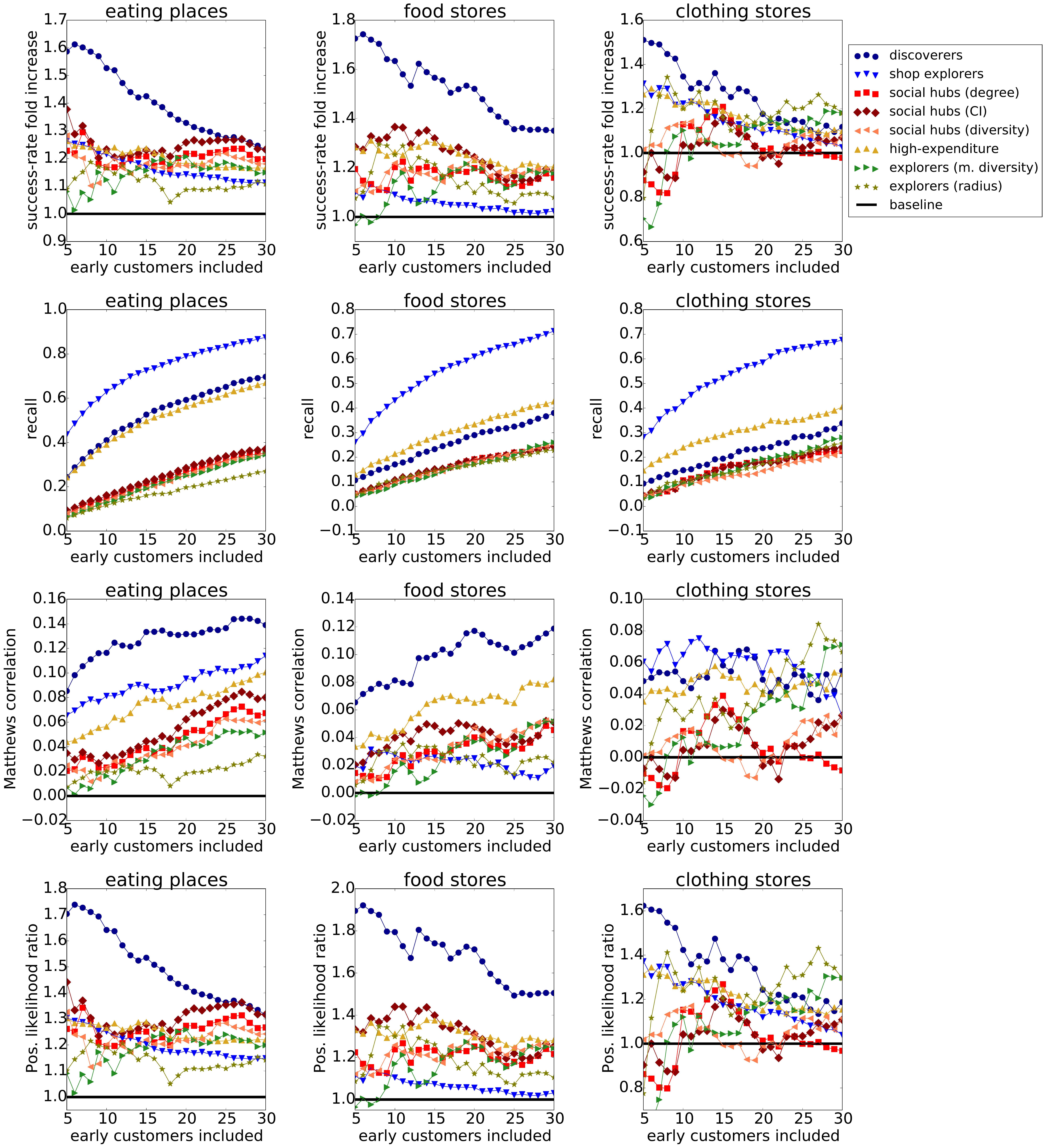}
   \caption{\textbf{Prediction results for a longer training period (from Dec. 2015 to July 2017, 20 months) and a longer validation period (from Aug. 2017 to May 2018, 10 months) compared to those used in the main text}. The other parameters of the analysis are the same as in the main text: $\Delta=90, z\%=10\%, s\%=10\%, \tau\%=1\%$.}
    \end{figure*}

\clearpage

\subsection*{Beyond classification: fold increase of the number of customers}

    \begin{figure*}[h]
\centering
   \includegraphics[scale=0.225]{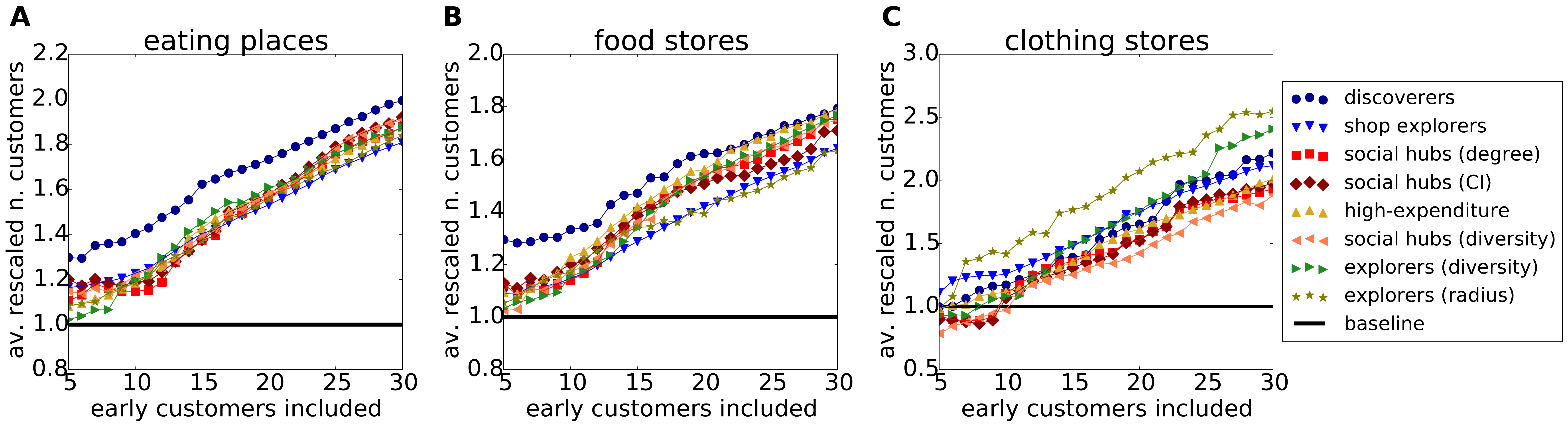}
   \caption{\textbf{Fold increase of the number of customers associated with early purchases by different groups of individuals}. For stores that received their first transaction within the validation period (excluding the last two months), we measure the ratio between their final number of first-time customers, $v_i$, and the average number of first-time customers of stores with the same MCC introduced in the same month. We refer to this ratio as the rescaled number of customers, $R(v)$. For each group of individuals, $\mathcal{I}$, we measure the average rescaled number of customers of stores that received an individual in $\mathcal{I}$ among the earliest $w$ customers. This average can be interpreted as the fold increase of the number of customers associated with early purchases by individuals in $\mathcal{I}$, compared with stores of the same age and category. We find that the discoverers exhibit the largest fold increases for eating places and food stores, whereas explorers by radius of gyration exhibit the largest fold increase for clothing stores. The parameters for the discoverer training are the same used in the main text: $\Delta=90, z\%=10\%, \tau\%=1\%$.}
    \end{figure*}

\end{document}